\documentclass[a4paper,11pt]{article}
\pdfoutput=1 

\usepackage{jheppub} 

\usepackage{bm}
\usepackage{tikz}
\usepackage{multirow}
\usepackage{amsmath}
\usepackage{slashed}
\usepackage{braket}
\usepackage{amstext,amssymb}
\usepackage{graphicx}
\usepackage{booktabs}
\usepackage{xspace}
\usepackage{color}
\usepackage{units}
\usepackage[normalem]{ulem}

\newcommand{\mathsym}[1]{{}}

\newcommand{\hs}{\hspace*{0.02 true cm}}

\newcommand{\bav}{\begin{array}{cccc}}

\usepackage{wrapfig}
\definecolor{light-gray}{gray}{0.95}
\usepackage{tcolorbox}
\usepackage{authblk}

\begin{document} 
\begin{center}
{\Large\bf \hspace{-0.5cm} Concurrence fill and mode distribution of entanglement in neutrino oscillation}
\end{center}

\vspace{0.1cm}

\begin{center}

 {\hspace{-1.5cm}Rajrupa Banerjee$^\dagger$}~\footnote{rajrupab@iitbhilai.ac.in}, {Prasanta K. Panigrahi$^{\ddagger,\star}$}~\footnote{panigrahi.iiser@gmail.com}, {Hiranmaya Mishra{$^{\star\star,\S}$}}~\footnote{hiranmaya@niser.ac.in}, {Sudhanwa Patra$^{\dagger,\S}$}~\footnote{sudhanwa@iitbhilai.ac.in}
\\
\vspace{0.2cm}
{\it  $^\dagger$ Department of Physics, Indian Institute of Technology Bhilai, Durg-491002, India\\
\it \hspace{-0.5cm}$^\ddagger$Center for Quantum Science and Technology, SOA University, Bhubaneswar-751030, India\\
\it \hspace{-0.5cm}$^\star$Indian Institute of Science Education and Research Kolkata, Mohanpur-741246, India\\
\it \hspace{-0.5cm}$^{\star\star}$ School of Physics, National Institute of Science Education and Research,
\\
An OCC of 
Homi
Bhabha National Institute, Jatni- 752050, India\\
\it $^{\S}$ Institute of Physics, Sachivalaya Marg, Bhubaneswar-751005, India}
\end{center}
\begin{abstract}
\noindent 
The flavor oscillations in the neutrino system are known to be related to the multi-mode entanglement of a single particle state. 
Neutrino oscillations are shown to encompass not only quantum superposition among different mass eigenstates but also a single-particle multi-mode entanglement in the flavor basis as they propagate. In the framework of three flavor neutrino oscillations, we demonstrate that the measures of entanglement can be expressed in terms of experimentally accessible appearance and disappearance probabilities. We explicitly show here that the genuine tripartite entanglement measure, i.e., the tangle, vanishes identically for all flavors, signifying that a three flavor neutrino system forms a W-type entangled state. Further, we investigate alternative measures of tripartite entanglement like the partial tangle and the concurrence fill, which capture the total sharing of entanglement beyond pairwise correlations. In terms of bipartite and bipartitioned entanglement measures, we derive the symmetric invariant and the concurrence fill, which quantify the distributed entanglement and are completely expressible in terms of flavor transition probabilities. These entanglement measures display distinct energy dependent patterns across the oscillation window, which can be experimentally accessible in the long baseline experiments like DUNE, providing an alternative quantum information perspective on flavor evolution. 
We use General Long Baseline Experiment Simulator (\textsf{GLoBES}) simulations within the DUNE setup to investigate these tripartite entanglement measures in terms of neutrino energy and the length of the baseline. It is observed that, at the point of maximal mixing, these measures show near maximal entanglement between the muon and the tau flavor modes, establishing entanglement monogamy. Within the DUNE setup, the wide band of energy and expected higher sensitivity to CP violation at the second oscillation maximum provide a unique advantage to explore the quantum correlation effects across a broader energy window. 
Thus, the flavor coherence and CP-phase dependent interference can influence the quantum entanglement measures among neutrino flavor modes, and thereby, such potential new physics effects can be probed in current or future planned neutrino experiments.
\end{abstract}

\def\thefootnote{\arabic{footnote}}
\setcounter{footnote}{0}

\newpage

\section{Introduction}
\label{sec:sec-intro}
Neutrino oscillations provide a quantum mechanical description of flavor transitions among the three neutrino flavors $(\nu_e,\nu_\mu,\nu_\tau)$. In the standard three-flavor picture, a neutrino flavor state $\nu_\alpha$ $(\alpha = e, \mu, \tau)$ is expressed as a quantum superposition of the three mass eigenstates $\nu_i$ $(i=1,2,3)$. A neutrino is created in a definite flavor state. As the neutrino propagates, each mass eigenstate accumulates a different phase depending on the baseline and energy, and the interference among these phases gives rise to oscillations among different flavors over distance. 
This interference of mass components modulates the flavor content and generates the oscillation phenomena observed in accelerator, atmospheric, solar, and reactor experiments~\cite{Bilenky:2004xm,Kayser:2012ce,SNO:2002tuh,T2K:2011ypd,MINOS:2011amj,K2K:2006yov,DayaBay:2012fng,Super-Kamiokande:2002ujc,NOvA:2016vij}. The oscillation pattern is governed by the mass-squared differences $(\Delta m^2_{21},|\Delta m^2_{31}|)$, three mixing angles $(\theta_{12},\theta_{13},\theta_{23})$, the leptonic CP violating phase $\delta$, and the dynamical quantities $L$, $E$, and the matter potential $A = 2\sqrt{2}\,G_F N_e E$.

Quantum entanglement, on the other hand, is a fundamental nonclassical feature that arises directly from the superposition principle and plays a central role in quantum information science~\cite{Nielsen_Chuang_2010,PhysRevA.98.052351}. In controlled quantum systems, entanglement serves as a valuable resource that makes possible a variety of remarkable protocols—including quantum teleportation and quantum key distribution~\cite{PhysRevLett.70.1895,PhysRevLett.67.661}. Neutrino oscillations and quantum entanglement both arise from the principle of superposition, but they play fundamentally different roles. In oscillations, superposition governs how a single neutrino evolves among flavor states, while in entanglement, it describes inseparable correlations shared between different subsystems.
Superposition concerns coherent combinations within a single subsystem, while entanglement refers to non factorizable correlations between distinct subsystems.

Nevertheless, the three-mode structure of neutrino oscillations naturally raises the question of whether the evolving flavor state should be interpreted purely as a coherent superposition or as a form of single-particle multimode entanglement. This possibility has been explored in several works within the occupation-number representation, where the three flavor (or mass) modes define an effective tripartite system~\cite{Blasone:2007vw,Blasone:2010ta,Blasone:2013zaa,Blasone:2014jea,Blasone:2015lya,Banerjee:2015mha,Ettefaghi:2023zsh,Blasone:2022joq,Li:2022mus,Patwardhan:2021rej,Ettefaghi:2020otb}. However, a complete and experimentally relevant characterization of this single-particle multipartite entanglement, particularly in the context of realistic long baseline experiments, remains largely unexplored. This challenge is compounded by the fact that a neutrino continuously oscillates among separable and maximally entangled configurations as it evolves \cite{Banerjee:2024lih}. 

In the context of three flavor neutrino system, early theoretical studies demonstrated that flavor states, when formulated in the occupation number (mode) representation, intrinsically possess entanglement because the flavor content at any time is a coherent admixture of mass components~\cite{Blasone:2007vw,Blasone:2010ta,Blasone:2013zaa,Blasone:2014jea,Blasone:2015lya,Banerjee:2015mha,Jha:2020dav,Jha:2022yik,Cirigliano:2024pnm}. 
Follow up studies have shown that, once the three flavor states are treated within a common Hilbert space, both bipartite and tripartite forms of entanglement can arise during neutrino propagation. 
These analyses demonstrated that flavor oscillations continuously reorganize the underlying quantum correlations among the three modes, mirroring the correlation dynamics known from multimode quantum-information systems~\cite{Ettefaghi:2023zsh,Blasone:2022joq,Li:2022mus,Patwardhan:2021rej,Ettefaghi:2020otb}.

More recent investigations have extended this perspective to phenomenologically relevant settings, using entanglement based diagnostics to probe oscillation parameters, matter effects, and possible nonstandard interactions (NSI)~\cite{Banerjee:2024lih,Konwar:2024nrd,Konwar:2025ipv}. Complementary work on the quantum speed limit has further clarified the dynamical constraints governing flavor evolution~\cite{Bouri:2024kcl,Jha:2025ekn,Jha:2026jef}. 
 In addition, the role of entanglement among supernova neutrinos and its potential impact on observable signals at the DUNE experiment have been investigated in Ref.~\cite{Johny:2026obb}.  
 It may further be noted that quantum entanglement in the neutrino system has been studied from various complementary perspectives. In particular, in Ref.~\cite{Bittencourt:2023asd}, an analysis of tripartite entanglement in three flavor neutrino oscillation was carried out using complete complementary relations. It was explicitly demonstrated that, unlike pure bipartite neutrino states with vanishing local coherence, tripartite neutrino systems generally exhibit non-vanishing local coherences.
Extending these ideas, subsequently, a triality relation for three-flavor neutrino oscillations was explored in terms of predictability, visibility, and I-concurrence~\cite{Banerjee:2025vyh}. In parallel, a systematic study of quantum coherence in both two- and three-flavor neutrino oscillations has been performed, examining its energy and baseline dependence within the standard oscillation framework. This analysis complements the present entanglement results by isolating the role of coherence in governing flavor interference and CP-phase effects \cite{Alok:2025qqr}.

Together, these developments highlight that neutrino oscillations can be understood not only as flavor transitions but also as the evolution of a single-particle, multimode entangled state shaped by mixing dynamics and propagation effects.

It may be noted that the quantification of entanglement measures for multipartite systems is considerably more intricate than for bipartite ones~\cite{Mukherjee:2025edo}. Well established bipartite measures such as concurrence~\cite{Wootters:1997id,Hill:1997pfa}, negativity~\cite{Vidal:2002zz}, and entanglement of formation~\cite{Wootters:1997id}, do not trivially generalize to three mode systems.
Any tripartite system can form two distinct entanglement classes such as  $\ket{\text{GHZ}}=\frac{1}{\sqrt{2}}\left(\ket{000}+\ket{111}\right)$ and $\ket{\text{W}}=\frac{1}{\sqrt{3}}\left(\ket{100}+\ket{010}+\ket{001}\right)$ which differs in their global entanglement behaviour.
 For tripartite systems, these two inequivalent classes of genuine multipartite entanglement are distinguished by the tangle (or three tangle)~\cite{Coffman:1999jd}, which detects GHZ-type correlations but vanishes for W states. Additional tools such as partial tangles~\cite{Lee:2005qzt} help diagnose biseparability. More recently, the concurrence fill $C_F$~\cite{Li:2021epj} has been proposed as a geometric measure of genuine multipartite entanglement, based on the area enclosed by the squared concurrences \cite{Sarkar:2024myo}. This measure satisfies the essential axioms of non-negativity, monotonicity, discriminance, convexity, and smoothness, and is particularly sensitive to the distributed W class of entanglement \cite{PhysRevLett.127.040403, Aggarwal:2025fab}.

 Motivated by this growing theoretical foundation, in this work, we provide a comprehensive characterization of bipartite and genuine tripartite entanglement in the three-flavor neutrino system, expressing all measures directly in terms of experimentally measurable oscillation probabilities. 
A core objective of this work is to determine and to characterize in an experimentally meaningful way the specific class of quantum entanglement that emerges in the evolution of three flavor neutrino oscillations. Therefore, to make the global entanglement structure transparent, we have considered the W and GHZ states, which differ clearly in their physical properties.
We investigate the multipartite entanglement structure of the three flavor neutrino system using experimentally accessible quantities, i.e., appearance and disappearance probabilities in Deep Underground Neutrino Experiment (DUNE), thereby establishing a direct bridge between quantum information theory and long baseline 
neutrino phenomenology. We first construct a single particle, multimode description of an oscillating neutrino and show how the standard oscillation formalism emerges naturally from the entangled flavor state. We then introduce three complementary entanglement diagnostics tailored to the neutrino system: (i) tangle, (ii) partial tangles, and (iii) the concurrence fill. A key aspect of this work is that we provide explicit expressions for these measures directly in terms of neutrino oscillation amplitudes and measurable probabilities. We further perform detailed DUNE-based simulations to map these quantifiers as functions of energy and baseline, thereby identifying the dynamically accessible entanglement regimes.

Our results show that the tangle vanishes across the entire energy range, ruling out GHZ-type entanglement and suggesting separable, biseparable, or W-type structure. Computation of the pairwise concurrences reveals that entanglement is dominantly shared within the $\nu_\mu-\nu_\tau$ and $\nu_e-\nu_\tau$ pairs, with negligible correlations in the $\nu_\mu-\nu_e$ sector. Together with the partial-tangle analysis, this pattern provides clear evidence of biseparability in the tripartite neutrino state, both in matter and in vacuum. To further assess the genuine tripartite structure, we evaluate the concurrence fill and find that the area of the concurrence triangle varies with energy. A nonzero area indicates genuine three-mode entanglement, consistent with a W-type configuration, while the area vanishes at specific energies where tripartite correlations disappear. Analysis of the W-class inequality confirms that the neutrino system indeed occupies regions of W-type entanglement over significant portions of its evolution.

 In addition to characterizing the structure of flavor-mode entanglement, it is also important to examine whether such quantum correlations can provide practical insight into the fundamental parameters of the oscillation framework. In particular, the sensitivity to the Dirac CP phase $\delta_{CP}$ is one of the central goals of long-baseline experiments. Since the entanglement measures considered in this work are directly expressible in terms of oscillation probabilities, their variation with $\delta_{CP}$ carries potential information about CP violation. This motivates us to investigate not only the magnitude of entanglement but also its phase sensitivity in order to assess whether regions of enhanced entanglement response can serve as useful indicators for optimizing experimental measurements. Such a study of the correlation between $\delta_{\rm CP}$ and the entanglement measures is of particular relevance for the DUNE experiment, which covers a wide range of energies.

The structure of the paper is as follows. Section~\ref{sec:formalism} establishes the theoretical formalism and the mode-distribution picture of entanglement in neutrino oscillations. In Section~\ref{sec:measures}, we derive the multipartite entanglement measures, tangle, partial tangle, and concurrence fill, expressed directly in terms of reduced density matrices and oscillation probabilities. Section~\ref{sec:DUNE} details the DUNE simulation framework and presents numerical results, quantifying W-type entanglement as a function of energy and baseline under various mixing. In this section, we further analyze the sensitivity to the Dirac CP phase, focusing on the phase dependence of the concurrence fill $C_F$ as a potential probe of CP violation.
Finally, Section~\ref{sec:conclusion} summarizes our findings and discusses experimental prospects and future directions.
\section{Theoretical Framework}
\label{sec:formalism}
\label{theory}
The effective Hamiltonian in the flavor basis, for three flavor neutrino oscillation in the presence of matter, is expressed as~\cite{Pontecorvo:1957qd, Maki:1962mu, Pontecorvo:1967fh, Patra:2023ltl},
\begin{equation}
   \mathcal{H}=\frac{\Delta m_{31}^{2}}{2E}\big[U \hs\mbox{diag}(0,\alpha,1)\hs U^{\dagger}+\mbox{diag}(\Hat{A},0,0)\big],
\label{eq:effH}
\end{equation}
where $\alpha=\Delta m_{21}^{2}/\Delta m_{31}^{2}$ characterizes the ratio between solar and atmospheric mass square differences. The matter effect are encoded in $\Hat{A}=A/\Delta m^{2}_{31}$ with $A=2 V_{CC}$. The unitary mixing matrix diagonalizing the light neutrino masses turns out to be the same as well known Pontecorvo, Maki, Nakagawa, Sakata (PMNS) mixing matrix, i.e, $U_{\rm PMNS}\equiv U$
on the basis where the charged leptons are already diagonal. 
Using the standard parametrization in terms of three neutrino mixing angles $\theta_{ij}$ and a Dirac CP phase $\delta_{\rm CP}$, the form of the neutrino mixing matrix $U$ is given by
\begin{equation*} 
U\big(\theta, \delta_{\rm CP} \big) =\begin{pmatrix} 
 c_{12} c_{13} & s_{12} c_{13} & s_{13}e^{-i\delta_{CP}}\\
 -s_{12} c_{23} -c_{12} s_{13} s_{23}e^{i\delta_{CP}} & c_{12} c_{23} -s_{12} s_{13}s_{23}e^{i\delta_{CP}} & c_{13} s_{23}\\
 s_{13} s_{23}-c_{12} s_{13} c_{23}e^{i\delta_{CP}}   & -c_{12} s_{23} -s_{12} s_{13}c_{23}e^{i\delta_{CP}}  & c_{23} c_{13}  
 \end{pmatrix},
 \end{equation*}
 where, $c_{ij}= \cos{\theta_{ij}}$ and $s_{ij}=\sin{\theta_{ij}}$ ($i,j=1,2,3$). 

\noindent
For the case of three flavors in vacuum, the neutrino flavor states are defined as
\begin{equation}
|\nu_{\alpha} \rangle =\sum_{i=1,2,3} U\big(\theta, \delta_{\rm CP} \big)  |\nu_{i} \rangle
\label{eq:fltom}
\end{equation}
Here, $|\nu_{i} \rangle$'s are the mass eigenstates with definite energy $E_i$ with their propagation in vaccum described by 
$|\nu_{i}(t) \rangle = e^{-i E_{i} t} |\nu_{i} \rangle $. 
The time evolution of the flavor neutrino states is given by 
\begin{equation}
|\nu_{\alpha}(t) \rangle = \sum_{\beta=e,\mu,\tau}\,\widetilde{U}\big(t,\theta, \delta_{\rm CP} \big)_{\alpha \beta}  |\nu_{\beta} \rangle
\label{eq:tevolve}
\end{equation}
Using Eq.(\ref{eq:fltom}), the time evolution relation for the mass eigen states and reverting to the flavor basis, the evolution operator for the flavor basis $\widetilde{U}\big(t,\theta, \delta_{\rm CP}\big)$ is given by 
\begin{equation}
\widetilde{U}\big(t,\theta, \delta_{\rm CP} \big) 
= U\big(\theta, \delta_{\rm CP}\big)~ U_{0}(t)~ U\big(\theta, \delta_{\rm CP}\big)^{-1} 
\label{eq:ut}
\end{equation}
And $|\nu_{\beta}\rangle$ are the neutrino flavor states at $t=0$. 
Here, $U_{0}(t) = \mbox{diag}\big( e^{-i E_{1} t}, e^{-i E_{2} t}, e^{-i E_{3} t} \big) $ is the time evolution matrix for the mass eigenstates. Clearly, $\widetilde{U}\big(t=0,\theta, \delta_{\rm CP} \big) = 1$. Thus, the time evolved flavor state is given explicitly as, with $U \equiv U\big(\theta, \delta_{\rm CP}\big)$, 
\begin{eqnarray}
    \ket{\nu_{\alpha}(t)}&=&U_{\alpha 1} U^{*}_{e 1}e^{-i E_{1}t} \ket{\nu_{e}}+U_{\alpha 2}U^{*}_{\mu 2}e^{-i E_{2}t} \ket{\nu_{\mu}}   +U_{\alpha 3}U^{*}_{\tau 3}e^{-i E_{3}t} \ket{\nu_{\tau}},
    \label{eq:alphat}
\end{eqnarray}
The transition probability for $\nu_{\alpha} \to \nu_{\beta}$ is
\begin{equation}
P_{\alpha \beta} (t) = | \langle \nu_\beta \mid \nu_\alpha(t) \rangle |^2 = |\widetilde{U}_{\alpha \beta}(t)|^2\,.
\end{equation}
It is well known that, in the conventional approach of quantum mechanics, the neutrino oscillation is understood from the principle of superposition among three flavor states $\ket{\nu_e}$, $\ket{\nu_\mu}$, and $\ket{\nu_\tau}$. The quantum superposition in neutrino oscillation is based on the idea that an individual neutrino flavor state is expressed as a coherent linear superposition of the mass eigenstates $\ket{\nu_i}$ through the PMNS mixing matrix. Using evolution principles, the superposition characterizes the evolution of a single quantum degree of freedom, i.e, a neutrino flavor that changes continuously as the particle propagates. Thus, quantum coherence in oscillations reflects the wave nature of neutrinos without involving multipartite correlations among different flavors. As a result, the flavor amplitudes remain components of the single particle Hilbert space. 

In contrast to conventional two-particle entanglement, where correlations arise between physically distinct subsystems, single-particle entanglement (SPE) 
\footnote{The single particle multi-mode entanglement has already been investigated in Ref.~\cite{PhysRevA.72.064306}. The experimental verification along with the non-local correlation of single particle entanglement has also been established by \textit{Pasini et. al.}~\cite{PhysRevA.102.063708} in a photon system.}
refers to the distribution of quantum superposition across different modes of a single quantum system. In the neutrino context, the three flavor modes $(\nu_e,\,\nu_\mu,\,\nu_\tau)$ naturally define three orthogonal Hilbert-space sectors, each of which can be treated as a three level system in the occupation number basis,
\[
\mathcal{H}_i = \mathrm{span}\{0_i, 1_i\}, \qquad i\in\{e,\mu,\tau\},
\]
where $\ket{1_i}$ denotes the presence of a neutrino in flavor mode $i$ and $\ket{0_i}$ its absence.  
The whole Hilbert space is therefore the tensor product.
\[
\mathcal{H}=\mathcal{H}_e\otimes \mathcal{H}_\mu\otimes \mathcal{H}_\tau,
\]
with total dimension $2^3=8$. 
A flavor eigenstate such as $\ket{\nu_e}\equiv\ket{1_e,0_\mu,0_\tau}$ occupies only one flavor mode at production, but as the neutrino propagates, its state evolves into a coherent superposition distributed over all three modes. Formally, the whole Hilbert space of the three flavor modes  
\[
\big\{|000\rangle,\,|100\rangle,\,|010\rangle,\,|001\rangle,\,
|110\rangle,\,|101\rangle,\,|011\rangle,\,|111\rangle\big\}
\]
constitutes an orthonormal basis of the eight-dimensional occupation-number space. However, the Pauli exclusion principle restricts a single neutrino to the single-excitation subspace,
\begin{equation}
\mathcal{H}_{\rm 3Flav}
=\mathrm{span}\{|100\rangle,\;|010\rangle,\;|001\rangle\}, 
\qquad 
\mathrm{dim}(\mathcal{H}_{\rm 3Flav})=3,
\end{equation}
which is precisely the physically relevant sector for three-flavor oscillations.

Tracing out mode subspaces reduces the full $8\times 8$ density matrix to an effective $3\times 3$ object. This reduced description naturally aligns with the standard three-flavor oscillation formalism and provides a clean algebraic framework for analyzing the coherence and entanglement distribution among $\nu_e$, $\nu_\mu$, and $\nu_\tau$.
Therefore, within this single excitation subspace, the time-evolved flavor state takes the form  
\begin{align}
\ket{\nu_\alpha(t)}
    &=\,\widetilde{U}_{\alpha e}(t)\ket{1_e,0_\mu,0_\tau}
      +\widetilde{U}_{\alpha\mu}(t)\ket{0_e,1_\mu,0_\tau}
      +\widetilde{U}_{\alpha\tau}(t)\ket{0_e,0_\mu,1_\tau},
\label{eq:alphat-ent}
\end{align}
with normalization $\sum_{\beta}|\widetilde{U}_{\alpha\beta}(t)|^2=1$.
The quantities $\widetilde{U}_{\alpha\beta}(t)$ serve as the matrix elements of the flavor evolution operator, carrying the effects of mixing and propagation while fitting naturally into the three-mode tensor-product description. 
This mapping establishes the explicit correspondence
\begin{align}
\ket{\nu_e}   &= |1_e,0_\mu,0_\tau\rangle, \quad \ket{\nu_\mu} = |0_e,1_\mu,0_\tau\rangle, \quad \ket{\nu_\tau} = |0_e,0_\mu,1_\tau\rangle~,
\label{eq:ent-states-new}
\end{align}
showing that the three flavor neutrino system is isomorphic to a three qubit single excitation state. 
As the neutrino propagates, the relative phases acquired by $\widetilde{U}_{\alpha e}(t)$, $\widetilde{U}_{\alpha\mu}(t)$, and $\widetilde{U}_{\alpha\tau}(t)$, controlled by the solar and atmospheric mass splitting scales, continuously reshape the distribution of coherence among the modes. 
Neutrino oscillations, therefore, provide a natural physical realization of single particle multimode entanglement, where the particle number is fixed, and the occupation amplitudes of the three qubit modes evolve in time. This occupation number formulation, therefore, creates a direct conceptual and algebraic bridge between neutrino oscillation theory and quantum information science --- the system behaves as a controlled three qubit evolution with naturally generated multimode entanglement.
The state $\ket{\nu_\alpha(t)}$ thus becomes an entangled superposition whose structure varies with both baseline and energy. 

In the next section, we show that key multipartite entanglement measures, tangle, partial tangle, concurrence, and concurrence fill, can be expressed entirely in terms of oscillation probabilities, making them directly accessible for experimental analyses.
\section{Entanglement Measures in terms of 
Oscillation Probabilities}
\label{sec:measures}
In the next step of our analysis, we turn to quantifying the various measures of entanglement that naturally arise in bipartite and tripartite systems relevant to three flavor neutrino oscillations. In many physical situations, oscillations effectively involve only two flavors. For instance, the $\nu_\mu \rightarrow \nu_\tau$ channel dominates in atmospheric neutrinos, while $\nu_e \rightarrow \nu_\mu$ governs solar neutrino transitions. For such two-flavor scenarios, several well established bipartite entanglement measures exist, including the von Neumann entropy, concurrence, and negativity. However, when extending from two to three flavors, the structure of quantum correlations becomes genuinely multipartite. In this regime, a richer set of entanglement diagnostics is required. Multipartite quantifiers such as the tangle, partial tangle, and concurrence fill offer a more complete geometric and algebraic description of how quantum correlations are shared among the three flavor modes. 
\begin{itemize}
    \item The tangle serves as a measure of genuine tripartite entanglement.
    \item The partial tangle captures how entanglement is distributed among different bipartitions.
    \item The concurrence fill quantifies the total amount of shared correlations beyond simple pairwise contributions.
\end{itemize}
 In the following discussion, we introduce these measures explicitly and express them in terms of experimentally accessible oscillation probabilities. To do so systematically, we employ a density matrix formalism, which provides a natural and transparent framework for deriving all entanglement observables used in the subsequent analysis.

The flavor state in Eq.~(\ref{eq:alphat-ent}) admits several natural bipartitions, allowing one to examine how quantum correlations are distributed among different subsets of flavor modes. To quantify these correlations, the essential tool is the reduced density matrix, obtained by tracing over one or more subsystems of the full state. Starting from the expression of the density operator of a specific flavor state, 
\[
\rho_{\alpha}(t)=\ket{\nu_{\alpha}(t)}\bra{\nu_{\alpha}(t)}\,.
\]  
One may trace out selected flavor modes to obtain the reduced description relevant for a given partition.
In this work, the primary focus is on the single flavor reduced density matrix, obtained by tracing over two of the three flavor modes. This quantity plays a central role because, experimentally, neutrino detection is flavor-selective, involving interactions with a single flavor at a time; therefore, the single mode density matrix contains all the information required to predict observable event rates. Furthermore, mostly the entanglement measures constructed from this reduced density matrix provide a direct measure of how strongly the chosen flavor is entangled with the remaining modes. To make this concrete, we consider an initial muon-neutrino state, $\nu_\mu$, relevant for accelerator-based long-baseline experiments such as DUNE. Under three-flavor mixing and propagation, the corresponding time-evolved density matrix takes the explicit form~\cite{Konwar:2024nrd, Li:2021epj}:
\begin{equation}
  \rho_{\mu}(t) =\begin{pmatrix}
          0 & 0 & 0 & 0 & 0 & 0 & 0 & 0\\
          0 & ~|\widetilde{U}_{\mu\tau}|^{2} ~ & ~\widetilde{U}_{\mu\tau}\widetilde{U}^{*}_{\mu\mu} ~&~ 0 ~&~ \widetilde{U}_{\mu\tau}\widetilde{U}^{*}_{\mu e} ~&~ 0 ~&~ 0 ~& 0\\
          0 &~ \widetilde{U}_{\mu\mu }\widetilde{U}^{*}_{\mu\tau} ~&~ |\widetilde{U}_{\mu\mu}|^{2} ~&~ 0 ~&~ \widetilde{U}_{\mu\mu}\widetilde{U}^{*}_{\mu e} ~&~ 0 & 0 & 0\\
          0 & 0 & 0 & 0 & 0 & 0 & 0 & 0\\
          0 & \widetilde{U}_{\mu e}\widetilde{U}^{*}_{\mu\tau} & \widetilde{U}_{\mu e}\widetilde{U}^{*}_{\mu\mu} & 0 & |\widetilde{U}_{\mu e}|^{2} & 0 & 0 & 0\\
          0 & 0 & 0 & 0 & 0 & 0 & 0 & 0\\
          0 & 0 & 0 & 0 & 0 & 0 & 0 & 0\\
          0 & 0 & 0 & 0 & 0 & 0 & 0 & 0\\
      \end{pmatrix}.\\
\label{eq:denm}
\end{equation}
Using this density‐matrix representation, the survival and transition probabilities follow directly as the diagonal elements of $\rho_{\mu}(t)$,
$$P(t)_{\nu_\alpha\rightarrow\nu_\beta} 
 \equiv |\rho_{\alpha} (t)|_{\beta \beta} 
 = |\Tilde{U}_{\alpha\beta}(t)|^2\,~, $$
which satisfies probability conservation identity $\sum_{\beta} P_{\alpha \to \beta} =1$. 
\subsection{Tangle and partial tangle in three-flavor neutrino oscillations}
We next intend to introduce different tripartite entanglement observables, such as the tangle and the partial tangle, which quantify the degree of pairwise correlations and their dynamical interplay within the single-particle multi mode framework. 
Within the framework of three flavor neutrino oscillations, with the oscillating neutrinos produced in a definite flavor state at the source, the temporal evolution yields a dynamical redistribution of quantum correlations among the three flavor modes. As a result, one can establish bipartite mode entanglement between the flavor subspaces as a direct consequence of neutrino mixing. This observation naturally motivates us to conduct a detailed, quantitative investigation of the entanglement sharability among different flavor modes during propagation.

These entanglement observables can be directly expressed in terms of the elements of the density matrix given in Eq.(\ref{eq:denm}). This links quantum information theoretic concepts to experimentally observable neutrino oscillation probabilities. The standard three flavor neutrino oscillation probabilities are given in terms of time dependent amplitudes $\widetilde{U}_{\alpha \beta}\big(t,\theta, \delta_{\rm CP} \big)$ as given in Eq.(\ref{eq:tevolve}) as,
\begin{eqnarray}
P_{\alpha e} = |\widetilde U_{\alpha e}|^2, \quad
P_{\alpha \mu} = |\widetilde U_{\alpha \mu}|^2, \quad
P_{\alpha \tau} = |\widetilde U_{\alpha \tau}|^2,
\end{eqnarray}
which satisfy the probability conservation condition
\begin{eqnarray}
P_{\alpha  e} + P_{\alpha  \mu} + P_{\alpha \tau} = 1.
\label{eq:probcons}
\end{eqnarray}
It is already stated in Eq.(\ref{eq:alphat-ent}) that the time evolved neutrino states form a single-particle three-mode 
state  corresponding to the propagation of an initial flavor eigenstate $\nu_\alpha$ 
where $\widetilde U_{\alpha \beta}$ denotes the time evolved amplitude for the transition 
$\nu_\alpha \to \nu_\beta$ which can be expressed in terms of baseline length and energy, including matter effects.
A natural way to quantify genuine tripartite entanglement in a three flavor neutrino system is through the tangle, $\tau_\alpha$, which measures the residual entanglement associated with a given flavor mode after subtracting all pairwise contributions. Within the three mode framework, the tangle for flavor $\alpha$ is defined as  
\begin{equation}
\tau_{\alpha} \;=\; C_{\alpha|\beta\gamma}^{\,2} \;-\; C_{\alpha\beta}^{\,2} \;-\; C_{\alpha\gamma}^{\,2},
\qquad (\alpha,\beta,\gamma = e,\mu,\tau)\,,
\label{eq:tangle-ABC}
\end{equation}
where $C_{\alpha|\beta\gamma}^{2}$ denotes the bipartitioned concurrence between flavor $\alpha$ and the composite subsystem formed by the remaining two modes, while $C_{\alpha\beta}^{2}$ and $C_{\alpha\gamma}^{2}$ represent the standard pairwise concurrences. This structure captures the distribution of the total quantum correlation associated with a particular mode between true tripartite entanglement and bipartite sharing.

For illustration, consider the accelerator based neutrino experiment where the initial state is $\nu_\mu$. Using the reduced density matrix of Eq.~(\ref{eq:denm}), the bipartitioned concurrence for the partition $\mu|(e\tau)$ can be expressed directly in terms of measurable oscillation probabilities as  
\begin{equation}
C_{\mu|(e\tau)}^{2} \;=\; 4\,P_{\mu\mu}\big(1 - P_{\mu\mu}\big)\,,
\label{eq:CABC-prob}
\end{equation}
where the dependence on $P_{\mu e}$ and $P_{\mu\tau}$ has been eliminated using probability conservation, Eq.~(\ref{eq:probcons}). Similarly, the square of the two-flavor concurrences takes the form~\cite{Blasone:2007vw}
\begin{equation}
C_{\mu e}^{2} = 4 P_{\mu e} P_{\mu\mu}, \qquad 
C_{\mu\tau}^{2} = 4 P_{\mu\tau} P_{\mu\mu}\,.
\label{eq:CAB-prob}
\end{equation}
Substituting Eqs.~(\ref{eq:CABC-prob}) and (\ref{eq:CAB-prob}) into the definition of the tangle in Eq.~(\ref{eq:tangle-ABC}) yields  
\begin{align}
\tau_{e}
&= 4\!\left[P_{\mu e} - \Big(P_{\mu e}^{2} + P_{\mu e}P_{\mu\mu} + P_{\mu e}P_{\mu\tau}\Big)\right] 
= 4 P_{\mu e}\,\Big[1 - (P_{\mu e} + P_{\mu\mu} + P_{\mu\tau})\Big] = 0~,
\nonumber \\
\label{eq:tau0}
\end{align}
where the final equality follows from probability conservation again. Identical arguments show that $\tau_\mu = \tau_\tau = 0$ as well.
Thus, the tangle vanishes identically for all flavors throughout the evolution. This result carries an important implication. The three flavor neutrino state cannot host GHZ-type entanglement (for which $\tau_\alpha$ is nonzero) but instead lies within the class of separable, biseparable, or W-type states. This motivates the need to examine additional multipartite diagnostics, such as partial tangles and concurrence fill, to distinguish among these possibilities and fully characterize the distribution of quantum correlations across the flavor modes.
\footnote{
The genuine tripartite entanglement $\tau$ and partial tangle for GHZ-type and W-class states are summarized as follows~\cite{Datta:2018fpp}:
\begin{eqnarray}
\boxed{
\begin{aligned}
\tau_\psi &\neq 0 \ \Longrightarrow\  \text{GHZ-type entanglement},\\
\tau_\psi &= 0,\ \tau_{ij}\neq 0 \ \Longrightarrow\  \text{W-type entanglement},\\
\tau_\psi &= 0,\ \tau_{ij}=0 \ \Longrightarrow\  \text{biseparable or separable state.}
\end{aligned}
}
\end{eqnarray}
}. 

Although the genuine tripartite entanglement (as quantified by the tangle) vanishes identically in the three flavor neutrino system, it remains meaningful to examine how entanglement is shared among pairs of flavor modes. A useful quantity for this purpose is the partial tangle $\tau_{\alpha\beta}$ between two modes $\alpha$ and $\beta$~\cite{Aggarwal:2025fab}, defined as  
\begin{equation}
\tau_{\alpha\beta}
    = \sqrt{C_{\alpha|\beta\gamma}^{2} - C_{\alpha\gamma}^{2}}
    = \sqrt{C_{\alpha|\beta\gamma}^{2} - C_{\alpha\beta}^{2} - C_{\alpha\gamma}^{2} + C_{\alpha\beta}^{2}}
    = \sqrt{\tau_\alpha + C_{\alpha\beta}^{2}}~,
\end{equation}
where the last equality follows from the definition of the tangle $\tau_\alpha$. 
Since the tangle satisfies $\tau_\alpha = 0$ for all flavors in the three-mode neutrino system, the partial tangle formally reduces to $\tau_{\alpha\beta} = C_{\alpha\beta}$ in situations where any one of the three transition probabilities vanishes, effectively collapsing the dynamics to a two-flavor sector. 
\begin{equation}
\tau_{e\mu} = 2\sqrt{P_{\alpha e}P_{\alpha\mu}} = C_{e\mu}, 
\qquad
\tau_{e\tau} = C_{e\tau}, \qquad
\tau_{\mu\tau} = C_{\mu\tau}.
\end{equation}
However, when all three probabilities $P_{\alpha e}$, $P_{\alpha\mu}$, and $P_{\alpha\tau}$ remain nonzero, the system genuinely explores the full three-flavor space, and in this regime the partial tangle no longer coincides with the pairwise concurrence, i.e., $\tau_{\alpha\beta} \neq C_{\alpha\beta}$. This distinction reflects the presence of nontrivial mode sharing across the three flavors and signals the onset of authentic three-mode entanglement redistribution.
Thus, within the neutrino oscillation framework, the partial tangles coincide exactly with the corresponding bipartite concurrences, implying that all nontrivial quantum correlations are distributed pairwise among the flavor modes rather than residing in a genuine tripartite component.

This equivalence between partial tangles and concurrences highlights that the entanglement structure of the three-flavor neutrino system is effectively bipartite, with correlations redistributed among different flavor pairs depending on energy and baseline. 
To capture the residual component of entanglement that cannot be attributed to any pairwise correlation alone, we now introduce the concurrence fill, a geometric measure that characterizes the genuinely tripartite contribution to the flavor-mode entanglement.
\subsection{Concurrence Fill in Terms of Neutrino Oscillation Probabilities}
Here, we introduce the concept of concurrence fill $C_F$, which 
offers an intuitive way to visualize and quantify how entanglement is shared among three flavor modes. Unlike the well-known tangle $\tau$, 
$C_F$ remains non-vanishing for both cases. 
Such a measure is more appropriate 
 
that reflects not only the presence of genuine tripartite entanglement but also quantifies how that entanglement is distributed among different flavor modes during evolution. 
\begin{figure}[htb!]
\centering
\vspace{-0cm}
\includegraphics[scale=0.40]{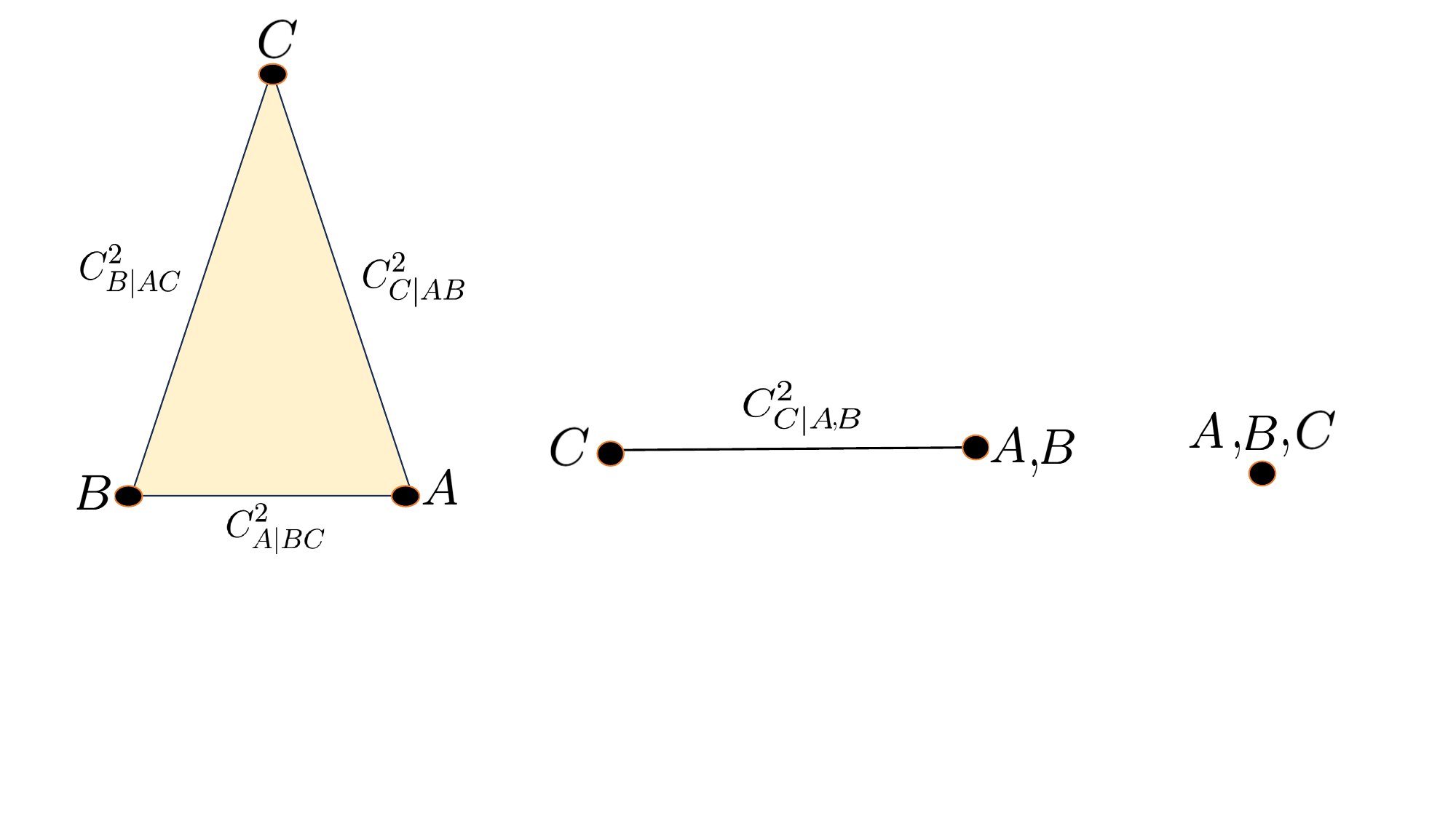}
\vspace*{-2cm}
\caption{Geometric illustration of the tripartite entanglement, biseparability, and fully separable (product) states. }
\label{fig:CF-area}
\end{figure}
To appreciate the concurrence fill, it is relevant to introduce a geometrical analogy using the triangle inequality. The geometric triangle inequality states that for any triangle, the sum of the lengths of any two sides must be greater than or equal to the length of the remaining side. Analogously, a similar relation, known as the entanglement monogamy inequality, has been established. This states that the bipartition entanglement of one subsystem and the rest cannot exceed the sum of the remaining bipartition entanglements in a tripartite system. In the experimental aspect of neutrino oscillation, this monogamy relation has already been explored in terms of CKW inequality \cite{Banerjee:2024lih}.
Motivated by this analogy, one can construct a concurrence triangle, where the sides represent the bipartition concurrences among the three partitions of a tripartite quantum system as shown in Fig.\ref{fig:CF-area}. 
The perimeter of this concurrence triangle quantifies the total tripartite entanglement shared among the three modes. The area of the concurrence triangle, termed as concurrence fill $C_F$, turns out to be a genuine measure of tripartite entanglement. 
\begin{table}[t]
\centering
\renewcommand{\arraystretch}{1.25}
\setlength{\tabcolsep}{6pt} 
\begin{tabular}{lccc}
\hline\hline
\textbf{States} & 
\textbf{\begin{tabular}[c]{@{}c@{}}W or GHZ state,\\ tripartite entanglement\end{tabular}} & 
\textbf{Pure biseparable state} & 
\textbf{Product state} \\
\hline
Perimeter     & $>0$ & $>0$ & $=0$ \\
Area          & $>0$ & $=0$ & $=0$ \\
Shortest edge & $>0$ & $=0$ & $=0$ \\
\hline\hline
\end{tabular}
\caption{Classification of tripartite entangled, biseparable, and product states in terms of geometric parameters.}
\label{tab:entanglement_class}
\end{table}
Using the formula for area of a triangle with sides $C^2_{A|BC}$, $C^2_{B|CA}$ and $C^2_{C|AB}$, the square of the concurrence fill $C^2_F$ which is the area can be expressed as
\begin{equation}
C^2_F = 
\left[
\frac{16}{3}\, Q(Q - C_{A|BC}^2)(Q - C_{B|AC}^2)(Q - C_{C|AB}^2)
\right]^{1/2},
\label{eq:CF_neutrino}
\end{equation}
where
\begin{equation}
Q = \frac{1}{2}\big(C_{A|BC}^2 + C_{B|AC}^2 + C_{C|AB}^2\big)~,
\end{equation}
represents the semiperimeter, analogous to the global entanglement of the system. The extra prefactor $16/3$ appears to ensure proper normalization, such that $0 < C_F < 1$.

In the context of the concurrence triangle as shown in figure \ref{fig:CF-area}, when the three sides of the triangle are comparable enclosing a finite area, this finite area signifies genuine tripartite entanglement among different modes.  
If the length of the shortest side of the concurrence triangle becomes zero, the triangle degenerates into a straight line, corresponding to a completely biseparable state. Furthermore, if any of the three sides vanish, the system collapses into a fully separable state corresponding to the absence of any entanglement between the different modes.
Thus, the concurrence fill $C_F$ serves not only as a quantitative measure of the amount of entanglement, but also as an indicator of how that entanglement is distributed among the three flavor modes.  
The concurrence fill attains the
fixed value $C_F^2 = 8/9$ for the symmetric $W$ state, $\ket{W}= \frac{1}{\sqrt{3}}\left(\ket{100}+\ket{010}+\ket{001}\right)$
and exhibits an upper limit of $C_F=1$ for a maximally tripartite entangled state, i.e., GHZ state.  
This contrast highlights the role of $C_F$
 as a geometric indicator capable of distinguishing between inequivalent classes of three-qubit entanglement.
In the context of neutrino oscillations, $C_F$ is not constant. Instead, it varies continuously with the oscillation parameters. In such a system, the evolution of $C_F(L, E)$ as a function of baseline and energy provides a direct means of tracing the redistribution of flavor coherence and mode entanglement among the three flavor sectors.
Using the relation for bipartitioned concurrence square $C_{\alpha|{\beta\gamma}}^2(L,E)$ given in Eq.(\ref{eq:CABC-prob}), the symmetric factor $Q$~\cite{Coffman_2000} is expressed in terms of oscillation probabilities as, 
\begin{equation}
Q  = 2\Big(P_{\alpha e}P_{\alpha \mu} + P_{\alpha \mu}P_{\alpha \tau} + P_{\alpha \tau}P_{\alpha e}\Big).
\label{eq:Q-prob-general}
\end{equation}
Thus, $Q$ characterizes the overall distribution of bipartitioned concurrences among all flavor modes. 
Using the above Eq.(\ref{eq:Q-prob-general}), the concurrence fill $C^2_F$ of Eq.(\ref{eq:CF_neutrino}) can be explicitly written in terms of oscillation probabilities as,
\begin{equation}
\boxed{
C_F(\nu_\alpha;L,E) =
\left[
\frac{16}{3}\;
Q
\prod_{\beta=e,\mu,\tau}
\Big(Q - 4\,P_{\alpha\beta}\,(1 - P_{\alpha\beta})\Big)
\right]^{1/4}\,.
}
\label{eq:CF_general_prob}
\end{equation}
The equations (\ref{eq:Q-prob-general}) and (\ref{eq:CF_general_prob}) 
In the following section, we analyze the tangle, partial tangles, and concurrence fill $C_F$ within the DUNE experimental configuration, demonstrating that the experiment’s precise measurements of the oscillation probabilities $P_{\alpha\beta}(L, E)$ provide a direct and experimentally realizable route to verifying and reconstructing the underlying pattern of bipartite and genuine tripartite entanglement shared among the flavor modes.
\section{Entanglement Analysis at DUNE using GLoBES}
\label{sec:DUNE}
It is useful to emphasize that the DUNE experimental configuration provides an exceptionally well-suited environment for numerically investigating the tripartite entanglement structure of three-flavor neutrino oscillations. In particular, the long baseline of $1300~\mathrm{km}$, together with the broad-band neutrino beam spanning $E\simeq 1$--$5~\mathrm{GeV}$, enables DUNE to probe the full three-flavor oscillation pattern, including matter-induced modifications and multiple oscillation maxima. 
This broad energy reach is especially advantageous for evaluating quantum-informational quantities such as bipartite and bi-partitioned concurrences, the tangle, partial tangle, and the concurrence fill.

The characteristic features of a long-baseline $\nu_{\mu}\rightarrow\nu_{e}$ appearance measurement are governed by the ratio $1.27\,\Delta m^{2}_{31}\,L/E$. From the leading oscillatory term, the positions of the oscillation maxima satisfy
\begin{eqnarray}
    \frac{L}{E} \sim (2n-1)\times 500~\mathrm{km/GeV},
\end{eqnarray}
where $n = 1,2,\ldots$ labels the first, second, and higher maxima. For the DUNE baseline of $L=1300~\mathrm{km}$, this results in oscillation maxima at energies,
\[
E_{I} \simeq 2.6~\mathrm{GeV}\quad (\mbox{1st Oscil. Maxima}),\quad  \qquad
E_{II} \simeq 0.86~\mathrm{GeV}\, \quad (\mbox{2nd Oscil. Maxima})\,.
\]
While all long baseline neutrino experiments can access the first oscillation maximum, the second oscillation maximum can be accessed by DUNE (and T2HK) due to the broad band in beam energy. 
Furthermore, the second oscillation maxima are particularly important because the interference term dependent on the CP violating phase $\delta$ is enhanced by a factor of $\sim 3$ relative to the first oscillation maxima, offering increased sensitivity to CP violating effects \cite{Ghosh:2014rna, Rout:2020emr}. Apart from accessing the second oscillation maximum, the flux factor for DUNE plays an important role in analyzing oscillation probabilities. 
The DUNE flux, distributed roughly over $0.5-3~\mathrm{GeV}$, therefore provides a unique opportunity to explore quantum-informational aspects of oscillations across different energy ranges, accommodating both first and second oscillation maxima. While the neutrino flux in most long-baseline experiments is narrowly peaked around the first oscillation maximum--exhibiting an approximately Gaussian profile shown in green-shaded region in Fig.~\ref{flux} (right-pannel)--the DUNE beam spans a substantially broader energy window displayed in yellow-shaded region (right-pannel). This wide-band coverage enables a detailed study of flavor evolution not only at the first oscillation maximum but also across several intermediate energy bins and, crucially, at the second oscillation maximum. As a result, DUNE offers an experimentally accessible pathway to probe how bipartite and tripartite entanglement measures evolve across multiple oscillatory regimes.

Since all entanglement measures discussed in section \ref{sec:measures} can be written directly in terms of the oscillation probabilities $P_{\alpha\to\beta}(L, E)$, DUNE offers direct and experimentally meaningful input for tracking the redistribution of quantum correlations among the flavor modes. The energy dependence of $P_{\mu e}$, $P_{\mu\mu}$, and $P_{\mu\tau}$ (as shown in left pannel of Fig.~\ref{flux}) can induce distinct and interpretable oscillatory patterns in different entanglement measures like the concurrence, the tangle, the partial tangles, and the concurrence fill. Motivated by these considerations, we next numerically examine the relevant tripartite entanglement measures throughout the DUNE energy window and their implications for manifestations of flavor-mode entanglements in realistic long-baseline settings.

\section*{\textrm{A.} Experimental and Simulation Details}

DUNE employs a $1300~\mathrm{km}$ baseline from the neutrino source at the Fermi National Accelerator Laboratory (FNAL) to the Sanford Underground Research Facility (SURF). To simulate the experiment, we use the configuration files provided in the DUNE Technical Design Report~\cite{DUNE:2021cuw}. These files correspond to an exposure of $624$~kt-MW-years, equivalent to 6.5~years of neutrino running and 6.5~years of antineutrino running, with a $40$~kt fiducial mass liquid-argon time-projection chamber (LArTPC) far detector and a $120$~GeV, $1.2$~MW beam.

This configuration corresponds to approximately ten years of data collection under the staging assumptions described in Ref.~\cite{DUNE:2020jqi}. All simulations in this work are performed using the General Long-Baseline Experiment Simulator (GLoBES) package~\cite{Huber:2004ka,Huber:2007ji}. The oscillation parameter values adopted in our calculations are listed in Table~\ref{osc11}.

\begin{table}[htb!]
    \centering
    \begin{tabular}{c c c c c}
    \hline\hline
      $\theta_{12}$   & $\theta_{13}$ & $\theta_{23}$ & $\Delta m_{21}^2~(\mathrm{eV}^2)$ & $\Delta m_{31}^2~(\mathrm{eV}^2)$ \\
      $33.68^{\circ}$ & $8.56^{\circ}$ &  $43.3^{\circ}$ & $7.49\times 10^{-5}$ & $\pm 2.513 \times 10^{-3}$ \\
    \hline\hline
    \end{tabular}
    \caption{Oscillation parameters used in the probability and entanglement calculations \cite{Esteban:2024eli}.}
    \label{osc11}
\end{table}

\subsection*{\textrm{B}. Bipartite and Bi-partitioned Concurrence}
With the simulation framework established and using the oscillation parameters given in Table~\ref{osc11}, we now proceed to the numerical evaluation of the entanglement measures introduced in section~\ref{sec:measures}. 
\begin{figure}[htb!]
\centering
\hspace{-1.2cm}
\includegraphics[scale=0.45]{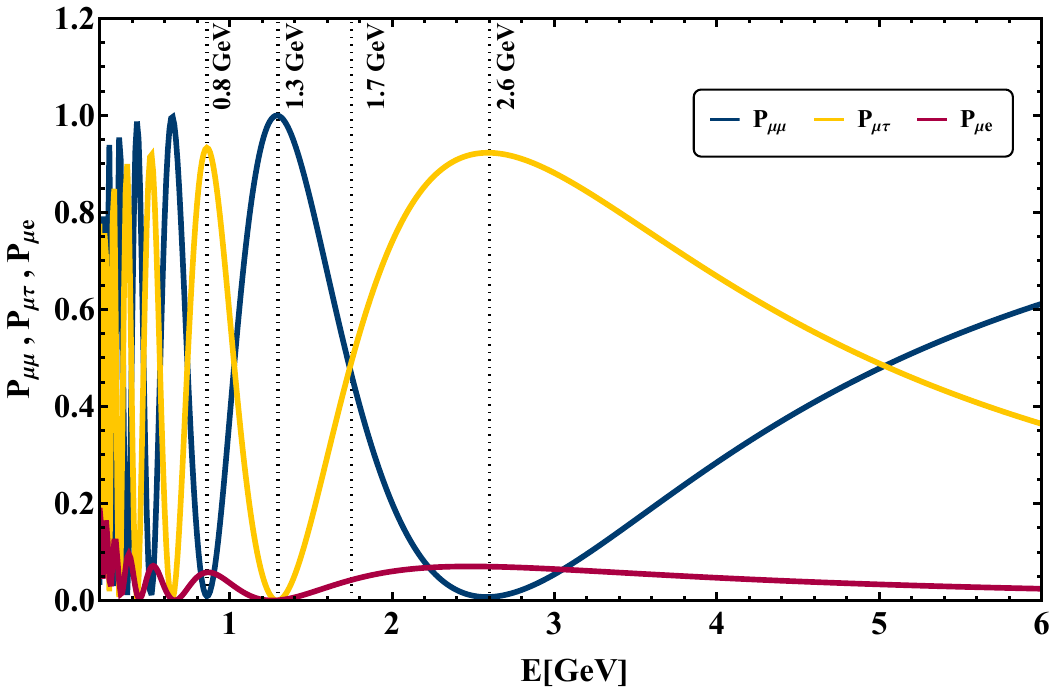}
\hspace{-0.2cm}
\includegraphics[scale=0.45]{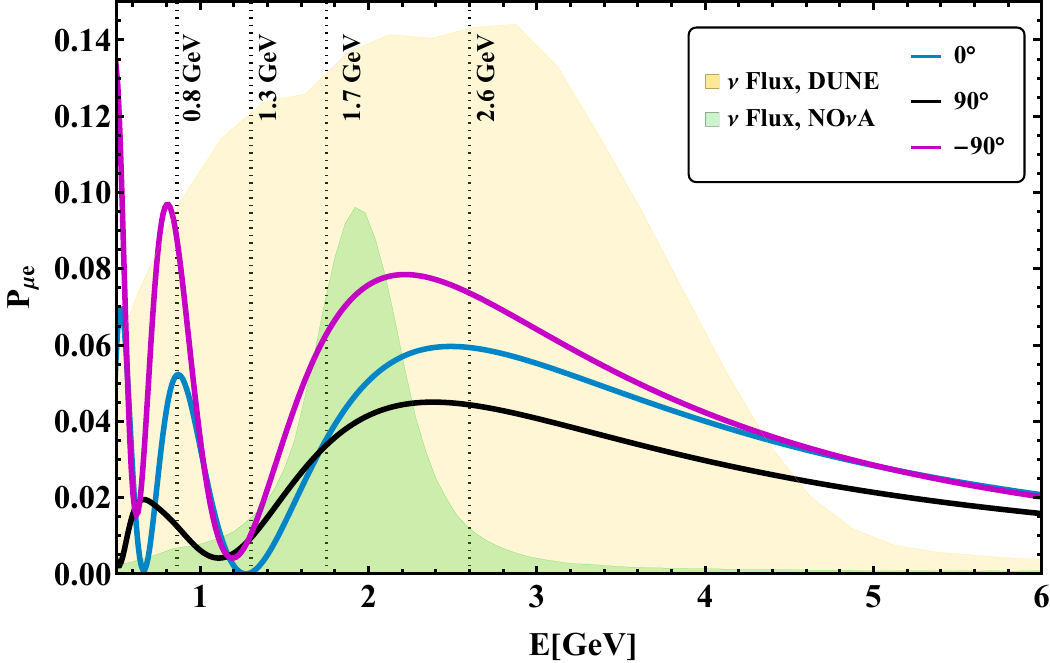}
\caption{Probability plots for $P_{\mu\mu}$, $P_{\mu e}$, and $P_{\mu\tau}$ (left panel) and the corresponding neutrino flux (shown up to a scale) alongside $P_{\mu e}$ (right panel) are presented for the DUNE experiment at three different values of the CP-violating phase $\delta$.}
\label{flux}
\end{figure}
Since DUNE is an accelerator-based experiment, the neutrino flux at production is $\nu_\mu$, the neutrino is created in a pure muon-flavor state, which in the occupation-number basis, is written as
\begin{equation}
\ket{\nu_{\mu}(t=0)} \equiv \ket{010}
= |0_e\rangle \otimes |1_\mu\rangle \otimes |0_\tau\rangle,
\end{equation}
corresponding to a single excitation in the $\mu$-mode and vacuum in the $e$- and $\tau$-modes.
This flavor eigenstate is a coherent superposition of three mass eigenstates, each of which during propagation acquires a dynamical phase $\Delta m_{\mathrm{sol}}^{2}L/(4E)$ or $\Delta m_{\mathrm{atm}}^{2}L/(4E)$. Combined with the PMNS mixing elements, the 
flavor state at a later time evolves to, as given in Eq.(\ref{eq:alphat-ent}), as
\begin{equation}
\ket{\nu_{\mu}(t)}
= \widetilde{U}_{\mu e}(t)\ket{100}
+ \widetilde{U}_{\mu\mu}(t)\ket{010}
+ \widetilde{U}_{\mu\tau}(t)\ket{001},
\end{equation}
The time-dependent amplitudes $\widetilde{U}_{\mu\alpha}(t)$ incorporate both mixing and propagation effects. Thus, these amplitudes directly determine the observable oscillation probabilities as, 
\[
P_{\mu\alpha}(L,E)=|\widetilde{U}_{\mu\alpha}(t)|^{2},
\]
\noindent
The behavior of the entanglement measures like bipartite and bi-partitioned concurrence, defined as in Eq.(\ref{eq:CABC-prob}) and Eq.(\ref{eq:CAB-prob}), is linked to the energy dependence of the probabilities. The numerical values of the survival and appearance probabilities, $P_{\mu\mu}$, $P_{\mu e}$, and $P_{\mu\tau}$, at selected energies are listed in Table~\ref{tab:prob}, where they explicitly satisfy probability conservation. This is reflected in the left panel of Fig.~\ref{osc11}, where we have plotted these probabilities as a function of energies. In Fig.~\ref{osc11}, we presented the evolution of oscillation probabilities $P_{\mu e}$ (red curve), $P_{\mu \mu}$ (blue curve), $P_{\mu \tau}$ (yellow curve) as a function neutrino energy which will be relevant for estimation of bipartite and bipartition concurrences. For example, the first oscillation peak for $P_{\mu \tau}$ occurs at $E\simeq 2.6$~GeV while the second oscillation peak occurs at $E\simeq 0.8$~GeV. The vertical dashed lines correspond to specific energies describing maximal mixing of $\nu_\mu$ and $\nu_\tau$ at $E\simeq 1.7$~GeV, no mixing at $E\simeq 1.3$~GeV, while the other two correspond to the first oscillation peak at $E\simeq 2.6$~GeV and the second oscillation peak at $E\simeq 0.8$~GeV.
The crossing of two probability curves $P_{\mu\mu}$ (blue) and $P_{\mu\tau}$ (yellow) at $E\simeq 1.7$~GeV correspond to maximal mixing between the flavors $\nu_\mu$ and $\nu_\tau$. 
This will help in understanding the effects of mixing and entanglement for specific energy ranges.  
The right panel Fig.~\ref{osc11} displays the corresponding appearance probability $P_{\mu e}$, clearly identifying both the first and second oscillation maxima that lie within the accessible DUNE energy window.
\begin{table*}[t]
\centering
\renewcommand{\arraystretch}{1.15} 
\setlength{\tabcolsep}{12pt} 
\begin{tabular}{c c c c}
\hline\hline
\textbf{Energy $E$ (GeV)}  &  $P_{\mu\mu}$  &  $P_{\mu\tau}$  &  $P_{\mu e}$  \\
\hline
2.6  & 0     & 0.92 & 0.08 \\
0.8  & 0     & 0.95 & 0.05 \\
1.3  & 1.00  & 0.00 & 0.00 \\
1.7  & 0.47  & 0.47 & 0.06 \\
\hline\hline
\end{tabular}
\caption{Oscillation probabilities $P_{\mu\mu}$, $P_{\mu e}$, and $P_{\mu\tau}$ at representative energies relevant for evaluating multimode entanglement in the three--flavor neutrino system. }
\label{tab:prob}
\end{table*}

For the present analysis, we focus on four representative neutrino energies that capture the essential features of flavor evolution across the DUNE spectrum. Two of these energies correspond to the first and second oscillation maxima, resulting in a large value of the appearance probability $P_{\mu e}$. The remaining two energies, 1.7~GeV and 1.3~GeV, lie between the two maxima and probe qualitatively different probability configurations. As seen in the left panel of Fig.~\ref{flux}, the point at 1.7~GeV satisfies the condition $P_{\mu\mu} \simeq P_{\mu \tau}$, marking a balanced regime in which disappearance and appearance channels contribute comparably. 

\begin{figure}[htb!]
\centering
\hspace{-1.2cm}
\includegraphics[scale=0.55]{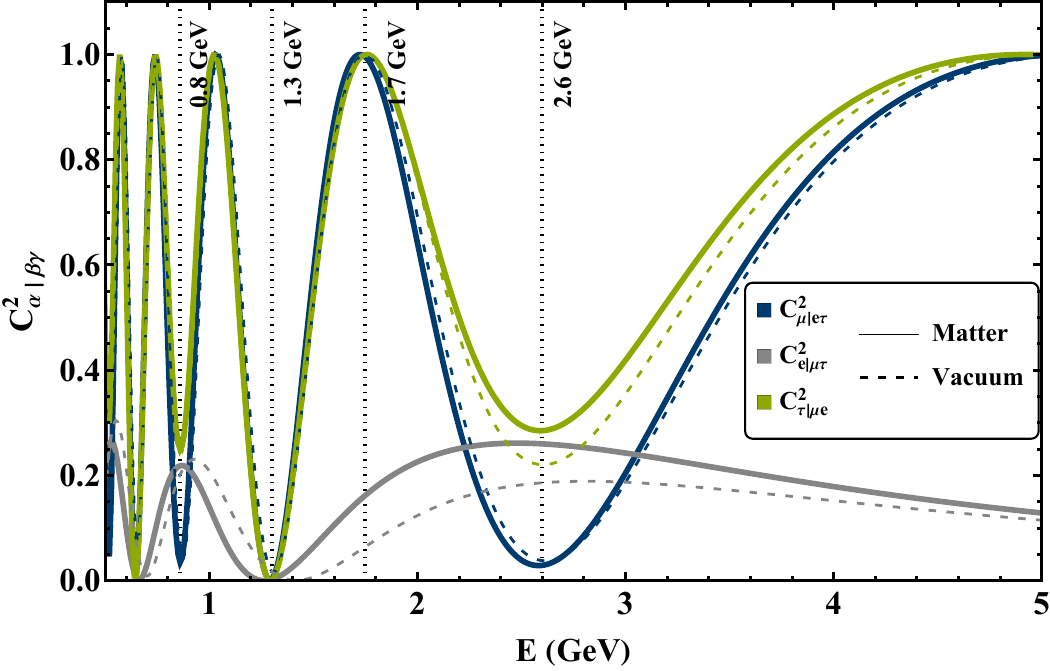}
\hspace{-0.2cm}
\caption{
 Bipartite entanglement structure of the three flavor neutrino system as a function of energy, shown for both vacuum (magenta) and matter (turquoise) propagation. The left panel displays the three bipartite concurrences $C^{2}_{\mu|e\tau}$ (solid green), $C^{2}_{e|\mu\tau}$ (dashed green), and $C^{2}_{\tau|\mu e}$ (dotted green), while the right panel shows the corresponding squared concurrences for the alternative bipartitions considered. Together, these curves map how quantum correlations are dynamically redistributed among the flavor modes across the full energy range relevant to long-baseline oscillations.
Vertical markers highlight representative energies corresponding to the first (2.6~GeV), intermediate (1.7~GeV and 1.3~GeV), and second (0.8~GeV) oscillation maxima. The behavior of the concurrences demonstrates that entanglement is not uniformly shared: different flavor pairs dominate at different energies, and matter effects further reshape this distribution by enhancing or suppressing selected bipartite correlations. Taken together, the two panels provide a complementary view of the flavor-mode entanglement landscape and illustrate how bipartite quantum correlations evolve under conditions relevant for experiments such as DUNE.
}
\label{concurrence}
\end{figure}
In contrast, at 1.3~GeV the probabilities satisfy $P_{\mu e}=P_{\mu\tau}=0$, so that $P_{\mu\mu}=1$, corresponding to a purely $\nu_\mu$ configuration with no flavor transition. 
These probability configurations establish the dynamical backdrop against which the multimode quantum correlations evolve.
Collectively, these four energies span the main dynamical regions of flavor evolution, offering natural benchmarks for studying how multimode entanglement varies across the DUNE spectrum. To explicitly examine this connection, we now evaluate bipartite entanglement measures across the same energy range. 
The two panels of Fig.~\ref{concurrence} provide a detailed picture of how bipartite entanglement evolves in a three-flavor neutrino system as it propagates in vacuum (dashed curves) and in matter (solid curves). The left panel shows the bipartitioned concurrences $C^{2}_{\mu|e\tau}$ (dark blue), $C^{2}_{e|\mu\tau}$ (gray), and $C^{2}_{\tau|\mu e}$ (green), each measuring how strongly one flavor mode is entangled with the remaining two. 

Let us recall that, for a flavor $\beta$, the bi-partitioned concurrence $C_{\beta|\mbox{rest}}^2=4P_{\alpha \beta} \big(1-P_{\alpha \beta}\big)$ with $\alpha$ being the initial flavor at $t=0$. With the initial muon flavor, let us analyze the energy dependence, for example, of $C^2_{\tau|\mu e}$ (green curve) at first oscillation maxima ($E\simeq 2.6$~GeV). At this point, $P_{\mu \tau} \simeq 0.92$ while $P_{\mu e}=0.08$ and $P_{\mu e}=0$ (see Table.\ref{tab:prob}). This makes $C^2_{\tau|\mu e}\simeq 0.28$ at this energy. 
Another interesting illustration can be observed near the first oscillation maximum, around $E \simeq 2.6~\mathrm{GeV}$ for the $1300~\mathrm{km}$ DUNE baseline. At this energy, the survival probability $P_{\mu\mu}$ nearly vanishes; correspondingly, the concurrence $C^{2}_{\mu|e\tau}$ drops to its minimum value ($\approx 0.03$), indicating that the $\nu_\mu$ mode becomes almost separable from the rest of the system. In contrast, the concurrence $C^{2}_{e|\mu\tau}$ reaches its largest value ($\approx 0.26$), reflecting the enhanced population of the $\nu_e$ and $\nu_\tau$ modes. Meanwhile, $C^{2}_{\tau|\mu e}$ remains comparably large ($\approx 0.28$), demonstrating that $\nu_e$ and $\nu_\tau$ retain significant quantum correlations even when the $\nu_\mu$ component is nearly absent.
A similar structure emerges at the second oscillation maximum, where the redistribution of flavor amplitudes again leads to enhanced correlations between the non-muon modes. In contrast, at $E = 1.3~\mathrm{GeV}$ all three concurrences vanish, signaling a fully separable configuration in which the state is purely $\nu_\mu$. At this energy, $P_{\mu\mu} \simeq 1$ while other probabilities $P_{\mu\tau}, P_{\mu e} \simeq 0$ leading to the vanishing of all three bi-partitioned concurrences $C^{2}_{\tau|\mu e}\simeq 0$, $C^{2}_{e|\mu \tau}\simeq 0$, $C^{2}_{\mu|e \tau}\simeq 0$
At $E = 1.7~\mathrm{GeV}$, however, the concurrences $C^{2}_{\mu|e\tau}$ and $C^{2}_{\tau|\mu e}$ reach their maximal value of unity, indicating that at this energy the entanglement is entirely shared between the $\nu_\mu$ and $\nu_\tau$ modes, with the electron flavor effectively decoupled.
The oscillatory rise and fall of these concurrences between $0$ and $1$ capture the essential quantum behavior of oscillations: the periodic creation, depletion, and reshuffling of entanglement among three coherently coupled modes.

Therefore, across the full DUNE energy band, the system cycles between weakly entangled and strongly entangled configurations, with different bipartitions becoming dominant at different energies. The strongest sharing of quantum correlations consistently occurs within the $\nu_\mu$--$\nu_\tau$ pair, as seen from the near overlap of the dark-blue and green curves. In comparison, the electron flavor plays a more modest role: its entanglement with $\nu_\tau$ remains noticeably smaller throughout the energy range, reflected in the lower amplitude of the gray curve of Fig.~\ref{concurrence}. Matter effects modify these patterns only mildly, confirming that the dominant structure of bipartite entanglement is set primarily by intrinsic mixing rather than by propagation in matter.

When these concurrence patterns (Fig.~\ref{concurrence}) are viewed alongside the probability curves (Fig.~\ref{flux}), a clear correlation emerges: enhancements in a particular transition probability strengthen the corresponding bipartitioned entanglement, while suppression reduces it. Taken together, the three concurrence curves offer a dynamic portrait of how flavor-probability oscillations translate into the exchange and redistribution of quantum information among the flavor modes.
This naturally motivates a closer examination of how these correlations are partitioned among flavor pairs, which we now pursue through the partial-tangle analysis.
\subsection*{\textrm{C}. Tangle (or Three-Tangle) and Partial Tangle}
\begin{figure}[htb!]
\centering
\hspace{-1.15cm}
\includegraphics[scale=0.55]{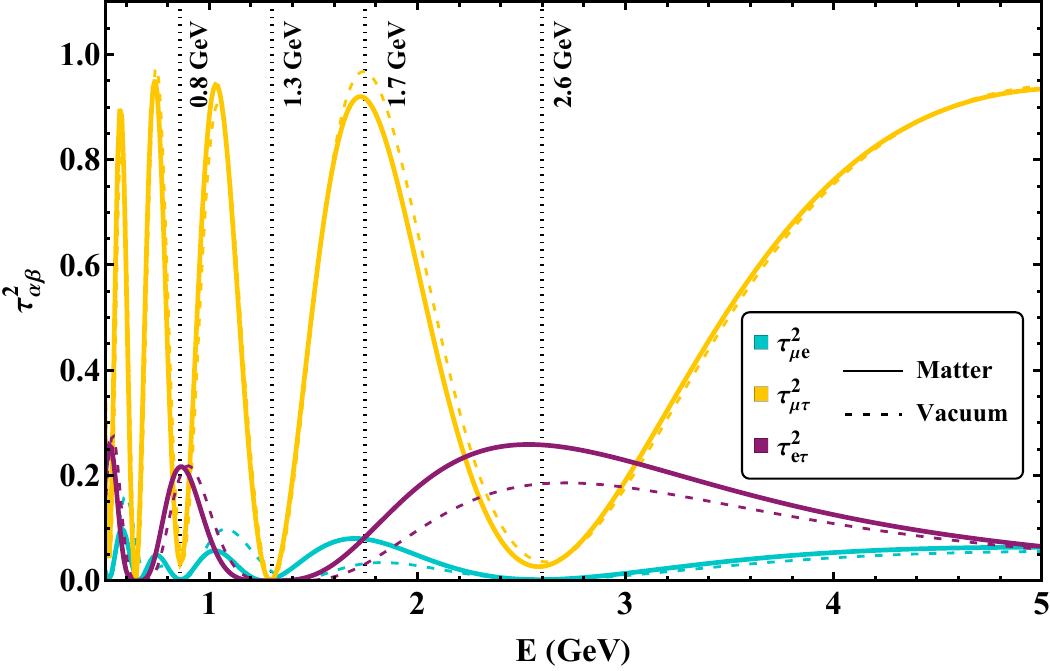}
\caption{
 Energy dependence of the partial tangles $\tau^{2}_{i|jk}$ in the three-flavor neutrino system for vacuum (dashed) and matter (solid) propagation. The curves correspond to the three bipartitions $\tau^{2}_{\mu e}$ (tuquoise), $\tau^{2}_{e\tau}$ (purple), and $\tau^{2}_{\mu\tau}$ (yellow), and quantify how the residual entanglement associated with each flavor mode is shared among the remaining two modes. Vertical markers indicate representative energies (2.6, 1.7, 1.3, and 0.8~\text{GeV}) corresponding to the first, intermediate, and second oscillation maxima. The variation of the partial tangles demonstrates that the distribution of residual entanglement evolves non-uniformly with energy. In vacuum, the structure reflects the coherent interplay among flavor components, whereas matter effects further reshape this pattern by enhancing or suppressing individual bipartitions. Overall, the figure provides a compact view of how tripartite entanglement dynamically reorganizes as the neutrino propagates, offering insight into the quantum structure relevant to long-baseline experiments such as DUNE.
}
\label{partialtangle}
\end{figure}
\begin{table*}[t]
\centering
\hspace*{-0.2cm}
\renewcommand{\arraystretch}{1.2} 
\setlength{\tabcolsep}{7pt} 
{\large
\begin{tabular}{c c c c c c c c c c c}
\hline
\hline
\textbf{E (GeV)} & $C^{2}_{\mu|e\tau}$ & $C^{2}_{e|\mu\tau}$ & $C^{2}_{\tau|\mu e}$ & $C^{2}_{\mu e}$ & $C^{2}_{\mu\tau}$  & $C^{2}_{e\tau}$ & $\tau^{2}_{e\mu}$ & $\tau^{2}_{e\tau}$ & $\tau^{2}_{\mu\tau}$ & $\tau_{\mu e\tau}$ \\
\hline
2.6 & 0.03  & 0.25 & 0.28 & 0 & 0.03 & 0.25 & 0 & 0.25 & 0.03 & 0\\
0.8 & 0.24  & 0.22 & 0.25 & 0 & 0.03 & 0.20 & 0 & 0.20 & 0.03 & 0\\
1.3 & 0 & 0 & 0 & 0 & 0 & 0 & 0 & 0 & 0 & 0 \\
1.7 & 1 & 0.16 & 1 & 0.1 & 0.1 & 0.25 & 0.08 & 0.08 & 0.92 & 0 \\
\hline
\hline
\end{tabular}
}
\caption{The table lists the squared bipartite concurrences for the three flavor bipartitions ($C^{2}_{\mu|e\tau}$, $C^{2}_{e|\mu\tau}$, $C^{2}_{\tau|\mu e}$) as well as for the two-mode reductions ($C^{2}_{\mu|e}$, $C^{2}_{\mu|\tau}$, $C^{2}_{e|\tau}$), providing a detailed view of how entanglement is distributed among different flavor pairs. The partial tangles ($\tau^{2}_{e\mu}$, $\tau^{2}_{e\tau}$, $\tau^{2}_{\mu\tau}$) quantify the residual correlations associated with each mode, while the tripartite tangle $\tau_{\mu e \tau}$ tests for GHZ-type entanglement. The selected energies (2.6, 1.7, 1.3, and 0.8~GeV) correspond to the first, intermediate, and second oscillation maxima. The pattern of values highlights pronounced reshuffling of bipartite correlations with energy, the emergence of residual entanglement in specific bipartitions, and the consistent absence of GHZ-type correlations ($\tau_{\mu e\tau}=0$). Overall, the table summarizes how the multimode entanglement structure of the three-flavor neutrino state evolves across characteristic oscillation regimes.}
\label{tab:tangle}
\end{table*}
\noindent
The genuine tripartite entanglement measure described by the tangle $\tau_{\alpha}$ as given in Eq.(\ref{eq:tangle-ABC}) vanishes for all flavors using the oscillation probabilities obtained from DUNE simulation as shown explicitly in Table.\ref{tab:tangle}.  
It is worthwhile to mention here that, in a three-mode quantum system, genuine tripartite entanglement can arise in two inequivalent forms--- GHZ-type,  
$\ket{\mathrm{GHZ}} = a\ket{000}+b\ket{111}$,  
and W-type,  
$\ket{\mathrm{W}} = x\ket{100}+y\ket{010}+z\ket{001}$ with $a,b,x,y,z$ are amplitudes for these basis states.  
The tangle is maximal for GHZ-type states but vanishes for all W-type, biseparable, and separable states. The identically zero value of $\tau_\alpha$ in neutrino oscillations therefore rules out GHZ-type correlations. It indicates that the system lies either in a biseparable or W-type entangled class, consistent with the concurrence patterns observed in Fig.~\ref{concurrence}. As a result of the vanishing of the tripartite entanglement measure, the tangle, for a neutrino system, is rather robust and arises from the conservation of oscillation probabilities. 

We next examine the partial tangle, $\tau_{\alpha\beta}$, which probes the residual bipartite entanglement between a selected pair $(\alpha,\beta)$ within a correlated tripartite system. Figure~\ref{partialtangle} displays the energy dependence of the three partial tangles.  
$\{\tau_{\mu e}^{2} \ \text{(cyan)},\ \tau_{\mu\tau}^{2} \ \text{(yellow)},\ \tau_{e\tau}^{2} \ \text{(purple)}\}$  
in both vacuum (dashed) and matter (solid) across the entire accessible energy range of DUNE. We observe that each partial tangle lies in the interval $[0,1]$ and exhibits pronounced oscillations with neutrino energy. 
\begin{itemize}
    \item At the first and second oscillation maxima ($E = 2.6~\mathrm{GeV}$ and $0.8~\mathrm{GeV}$), both $\tau^{2}_{\mu\tau}$ (yellow curve) and $\tau^{2}_{\mu e}$ (cyan curve) nearly vanish, indicating a strong suppression of residual entanglement involving the $\nu_\mu$ mode. In contrast, $\tau^{2}_{e\tau}$ (purple curve) becomes maximal (approximately $0.25$), showing that the dominant shareability of entanglement shifts to the $\nu_e$-$\nu_\tau$ pair.  
    \item At $E = 1.3~\mathrm{GeV}$ all partial tangles vanish simultaneously, reflecting a completely separable state, consistent with $P_{\mu\mu}=1$ and vanishing appearance probabilities of $\nu_e$ and $\nu_\tau$.
    \item  At $E = 1.7~\mathrm{GeV}$, however, all three partial tangles become nonzero:  
    $\tau_{\mu e}^{2}\approx \tau_{e\tau}^{2}\approx 0.08$ and $\tau^{2}_{\mu\tau}=1$. This pattern reveals that all three flavors participate in the entanglement, with the $\nu_\mu$-$\nu_\tau$ channel dominating, and equal nonzero values of $\tau_{\mu e}^{2}$ and $\tau_{e\tau}^{2}$ signaling the emergence of a fully developed W-type tripartite entangled state. 
\end{itemize}
Taken together, the concurrence and partial-tangle analyses build a unified and intuitive picture of how quantum correlations are distributed in the three-flavor neutrino system. Because partial tangles vanish for fully separable states and become simultaneously nonzero only for W-type states, they provide a clean diagnostic for identifying the underlying entanglement class. Situations in which one partial tangle is zero while the others remain nonzero correspond to biseparable configurations, where entanglement is shared only within a restricted flavor pair. The characteristic oscillatory patterns of $\tau_{\mu e}^{2}$, $\tau_{\mu\tau}^{2}$, and $\tau_{e\tau}^{2}$ therefore reveal the monogamy structure of the system: whenever entanglement strengthens in one flavor pair, it must diminish in the remaining two. From an experimental standpoint, the DUNE energy range of $0.8$--$3~\mathrm{GeV}$ coincides with the region where these redistributions of quantum correlations are most prominent. This makes DUNE particularly well suited for probing how entanglement flows among the flavor modes as functions of the oscillation phase $\Delta m_{ij}^{2}L/4E$ and the influence of matter effects.

This naturally motivates turning to the concurrence fill, a geometric measure of W-type entanglement, which provides an even more transparent visualization of how tripartite correlations evolve across the DUNE energy range.
\subsection*{\textrm{C}. Concurrence Fill}
\begin{figure*}[t]
\centering
\begin{minipage}[t]{0.49\textwidth}
  \centering
  \includegraphics[width=\linewidth]{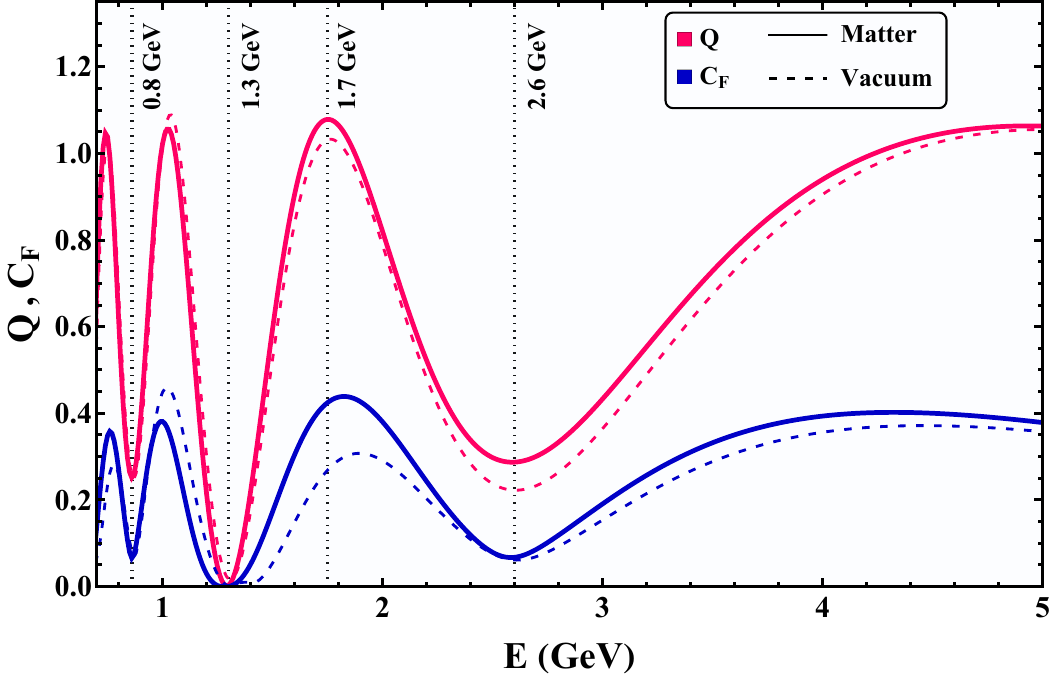}
\end{minipage}\hfill
\begin{minipage}[t]{0.49\textwidth}
  \centering
  \includegraphics[width=\linewidth]{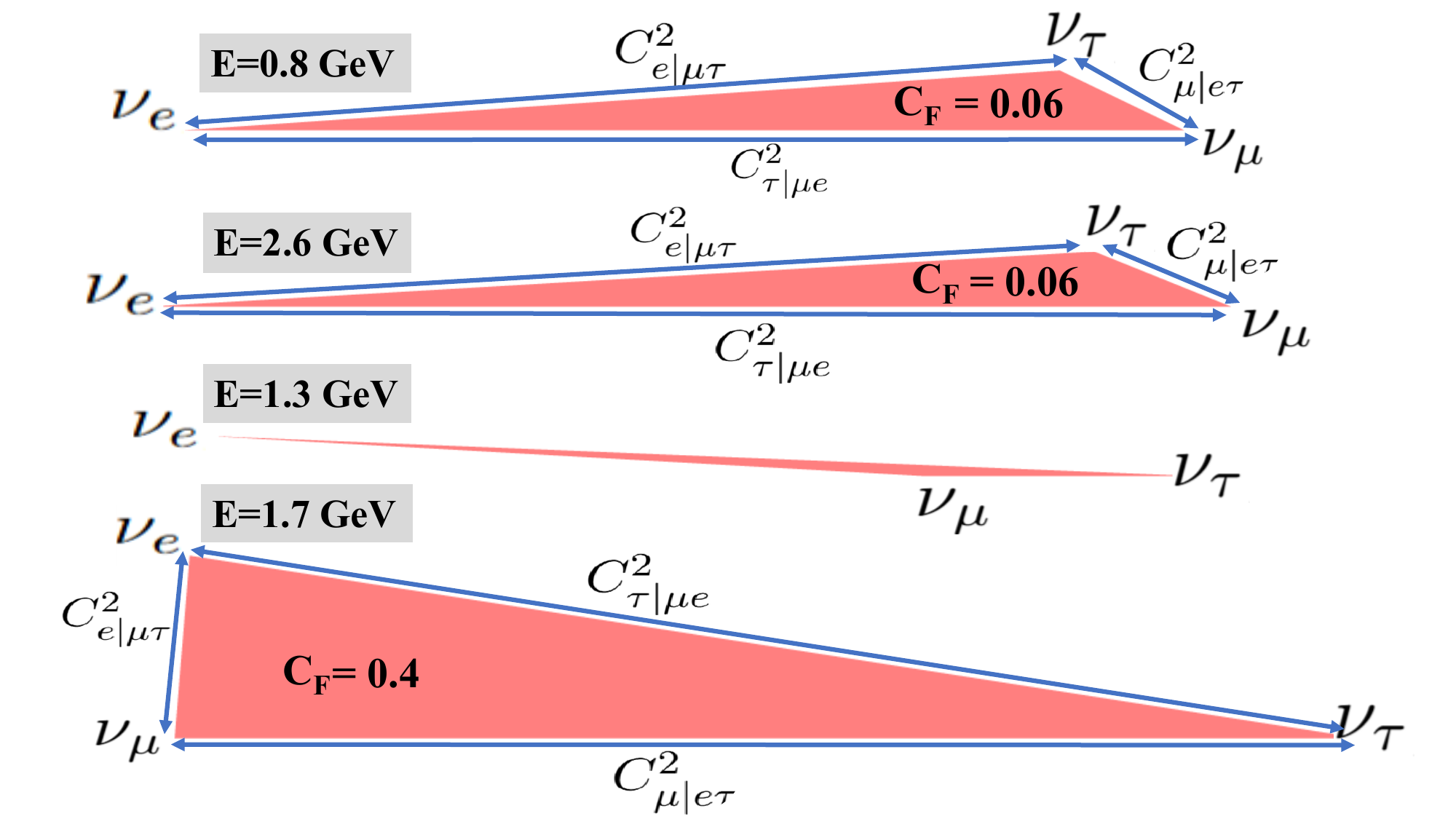}
\end{minipage}
\caption{
 Concurrence $Q$ (red), concurrence fill $C_F$ (blue), and concurrence triangles illustrating the energy dependence of bipartite and tripartite entanglement in the three flavor neutrino system. The left panel shows $Q$ and $C_F$ in the vacuum (dashed curves) and in matter (solid curves), with vertical markers indicating the first, intermediate, and second oscillation maxima. Matter effects modify the distribution of quantum correlations, leading to characteristic enhancements or suppressions relative to the vacuum. The right panel displays concurrence triangles constructed from $(C_{e\mu}, C_{e\tau}, C_{\mu\tau})$, where the enclosed area provides a geometric indicator of W type tripartite entanglement. Together, the panels summarize the dynamical redistribution of entanglement across energies relevant to long-baseline experiments such as DUNE.
}
\label{concurrencefill}
\end{figure*}

From the analysis of the partial tangle and concurrence, we find compelling evidence of bipartite entanglement embedded within the tripartite structure of neutrino oscillations. 
To achieve a more quantitative and geometrically intuitive characterisation of this behaviour, we now turn to the study of the concurrence fill, $C_F$, which serves as a robust and faithful geometric measure of the underlying entanglement structure of a qutrit system.  
\begin{table*}[t]
\centering
\renewcommand{\arraystretch}{1.0} 
\setlength{\tabcolsep}{8pt} 
{\large
\begin{tabular}{ c c c c c c c c c c}
\hline\hline
 $E$ (GeV) & $P_{\mu\mu}$ & $P_{\mu e}$ & $P_{\mu\tau}$ & $C^{2}_{\mu|e\tau}$ & $C^{2}_{e|\mu\tau}$ & $C^{2}_{\tau|\mu e}$ & Q & $C_F$ & $k^{2}$ \\
\hline
  2.6 & 0 & 0.92 & 0.06 & 0.03  & 0.26 & 0.28 & 0.30 & 0.07 & 0 \\
  0.8 & 0 & 0.95 & 0.05 & 0.46  & 0.18 & 0.59 & 0.25 & 0.06 & 0 \\
  1.3 & 1 & 0 & 0 & 0 & 0 & 0  & 0 & 0 & 0 \\
  1.7 & 0.47 & 0.47 & 0.04 & 1 & 0.15 & 1  & 1.08 & 0.44 & 0.01 \\
\hline\hline
\end{tabular}
}
\caption{
The table lists the survival and appearance probabilities ($P_{\mu\mu}, P_{\mu e}, P_{\mu\tau}$) together with the three bipartite concurrences ($C^{2}_{\mu|e\tau}$, $C^{2}_{e|\mu\tau}$, $C^{2}_{\tau|\mu e}$), concurrence triangle perimeter $Q$, the concurrence fill $C_F$ capturing genuine tripartite (W-type) correlations. The parameter $k$ enters the W-class inequality. These values illustrate how the pattern of bipartite and tripartite entanglement varies with energy: strong appearance probabilities correlate with enhanced bipartite concurrences, while nonzero $C_F$ identifies energy regions where genuine tripartite entanglement arises. The listed energies correspond to characteristic points in the oscillation spectrum, highlighting the dynamical redistribution of quantum correlations across different oscillation regimes.}
\label{tab:confill}
\end{table*}
In Fig.~\ref{concurrencefill}, the red and blue curves represent the half-perimeter ($Q$) and the concurrence fill ($C_F$), respectively. The dashed lines correspond to their evolution with neutrino energy in vacuum, while the solid lines represent the same quantities in the presence of matter effects. It is evident from the figure that the area of the concurrence triangle, quantified by $C_F$, varies with energy within the range $[0, 0.42]$. This variation implies that the concurrence area is not constant across the energy spectrum, reflecting a distinctly dynamical nature of neutrino entanglement. Specifically, over the energy interval $E \in [0.1, 5]~\mathrm{GeV}$, the system does not maintain a uniform degree of tripartite entanglement.  

To explore this behaviour in detail, we consider four benchmark energy values, as mentioned before. The corresponding concurrence triangles are shown in the left panel of Fig.~\ref{concurrencefill}. At the first and second oscillation maxima, the area of the triangle is minimal, with $C_F = 0.06$. At these energies, the survival probability $P_{\mu\mu}$ is nearly zero, while the appearance probabilities $P_{\mu e}$ and $P_{\mu\tau}$ reach their maxima (Fig.\ref{flux},Table~\ref{tab:prob}). Consequently, the entanglement between the $\mu$-$\tau$ flavor modes is strongly suppressed (Fig.~\ref{partialtangle}), represented by the shorter sides of the triangle (arms $A$ and $B$). In contrast, the entanglement between the appearance channels $e$-$\tau$ becomes dominant. The equal area of the concurrence triangles at these two energy points ($C_F = 0.06$) indicates that the degree of entanglement shared between the $e$ and $\tau$ flavors is comparable in both cases.  
At $E = 1.3~\mathrm{GeV}$, the concurrence fill $C_F$ vanishes entirely, corresponding to a situation where the concurrence triangle collapses into a straight line. This signifies the absence of genuine tripartite entanglement, marking a transition to a biseparable configuration. The area reaches its maximum value, $C_F \simeq 0.4$, at $E = 1.72~\mathrm{GeV}$, where both appearance and disappearance probabilities are nonzero. At this energy, the probabilities $P_{\mu\mu}$ and $P_{\mu\tau}$ become nearly equal, leading to two sides of comparable length, $C_{\mu|e\tau}$ and $C_{\tau|e\mu}$. The third side, associated with $C_{e|\mu\tau}$, also contributes significantly due to the finite value of $P_{\mu e}$.  
A quantitative comparison of the triangle areas, summarised in Table~\ref{tab:confill}, further substantiates this observation. The values $C_F(2.6~\mathrm{GeV}) = C_F(0.8~\mathrm{GeV}) = 0.06$ correspond to the minima, while $C_F(1.7~\mathrm{GeV}) = 0.4$ marks the maximum. Thus, $C_F(1.7~\mathrm{GeV}) > C_F(0.8~\mathrm{GeV}), C_F(2.5~\mathrm{GeV})$, indicating that the entanglement strength peaks at 1.7~GeV, where the interplay among all three flavor modes is most pronounced. The larger value of $C_F$ at this energy reflects a higher degree of genuine tripartite entanglement between the flavor modes within the neutrino system.  

The nonzero area of the concurrence triangle serves as a direct indicator of genuine tripartite entanglement in the three-flavor neutrino system. However, the mere presence of tripartite entanglement does not specify its class. The underlying state may belong either to the W family or to the GHZ family, which represents fundamentally different modes of multipartite correlation. To determine which class is realized during neutrino oscillations, it is essential to apply a more discriminating criterion. For this purpose, we now turn to an inequality-based analysis that provides a quantitative method for distinguishing W- from GHZ-type or biseparable configurations.
\subsection*{\textrm{D}. W class inequality.}
\begin{figure}[htb!]
\centering
\hspace{-1.2 cm}
\includegraphics[scale=0.55]{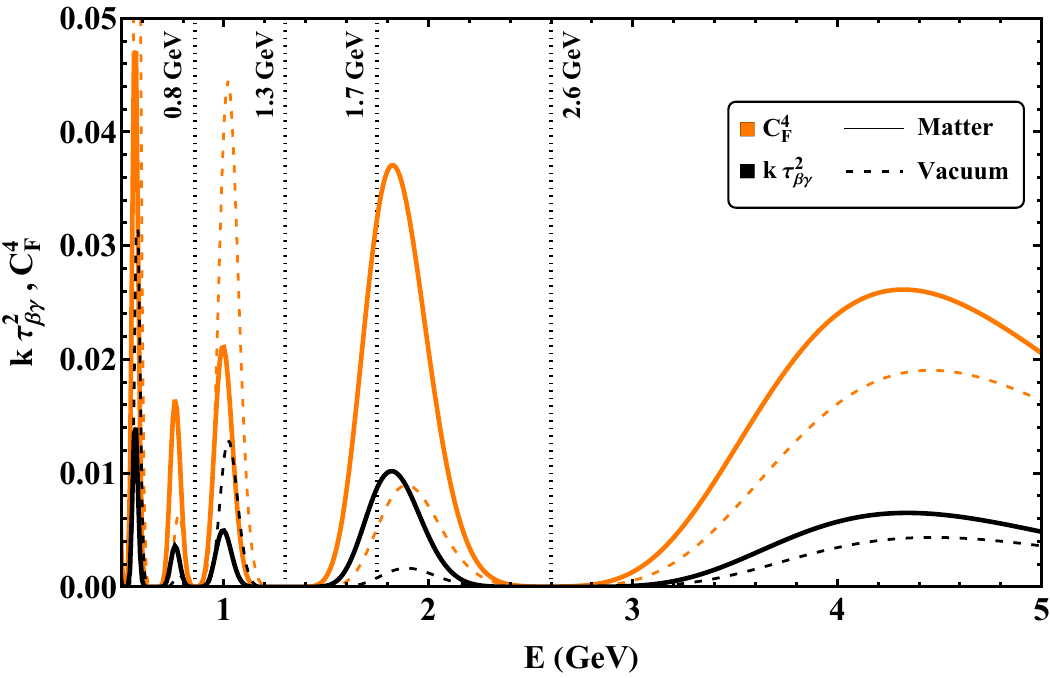}
\caption{Energy dependence of the W-class entanglement indicators for the three-flavor neutrino system in vacuum and matter. The upper panel shows the behavior of the concurrence fill $C_{F}^{4}$, while the lower panel displays the quantity $k\,\tau^{2}_{BC}$, which together enter the inequality used to determine whether the tripartite neutrino state belongs to the W class. The marked energies (2.6, 1.7, 1.3, and 0.8~GeV) correspond to the first, intermediate, and second oscillation maxima relevant for long-baseline experiments. The comparison between vacuum and matter illustrates how the presence of matter modifies the strength of multipartite correlations, with $C_{F}^{4}$ and $k\,\tau^{2}_{BC}$ showing distinct energy-dependent enhancements or suppressions. Regions where $C_{F}^{4}$ exceeds $k\,\tau^{2}_{BC}$ indicate that the W-class condition is satisfied, providing direct evidence for genuine tripartite entanglement within specific energy ranges. Overall, the figure highlights how the accessibility of W-type correlations evolves dynamically with energy and demonstrates the sensitivity of multipartite entanglement to matter effects in oscillation environments relevant for DUNE.
}
\label{inequality}
\end{figure}
From the above analysis, we find that for most of the chosen benchmark energies, the neutrino state exhibits genuine tripartite entanglement. However, at one particular benchmark point, $E = 1.3\,\mathrm{GeV}$, the state approaches a nearly biseparable configuration, indicating 
a qualitative change in the underlying entanglement structure. To identify which class the neutrino system corresponds to, we make use of the following inequality, which serves as a distinguishing criterion for W-type entanglement:
\begin{equation}
    \left[C_{F}\left(\left|\psi_{\mu e \tau}\right\rangle\right)\right]^{4} \geq k\,\tau_{e\tau}^{2},
\label{eq:inequality}
\end{equation}
where
\[
k = 2\left(\tau_{\mu e}^{2} + C_{\mu e}^{2}\right)^{4/3}
      \left(\tau_{\mu\tau}^{2} + C_{\mu\tau}^{2}\right)^{4/3}
      \left(\tau_{e\tau}^{2} + C_{e\tau}^{2}\right)^{1/3}.
\]
In Fig.~\ref{inequality}, the orange and black curves correspond to $C_{F}^{4}$ and $k\tau_{e\tau}^{2}$, respectively. The plot shows that this inequality is satisfied throughout the entire energy range considered, confirming that the neutrino system consistently exhibits the characteristics of a W-type state. The relation becomes an equality near the first and second oscillation maxima, where the area of the concurrence triangle reaches its minimum. This indicates that the tripartite correlations are weakened at these points, and the system temporarily behaves more like a collection of bipartite subsystems. 

At $E \simeq 1.7~\mathrm{GeV}$, where the concurrence area is largest, the inequality is strongly satisfied, signaling a robust W-type entanglement among all three flavor modes. Interestingly, this condition continues to hold over a broad energy window, roughly between $3$ and $8~\mathrm{GeV}$, suggesting that higher-energy neutrinos tend to maintain W-type correlations more dominantly. Overall, the inequality analysis 
provides a clear and quantitative confirmation that the entanglement structure in the three-flavor neutrino system belongs primarily to the W class, rather than the GHZ class of tripartite states.
\subsection*{E. Entanglement oscillograms and CP-phase sensitivity}
\begin{figure}[htb!]
\centering
\includegraphics[scale=0.55]{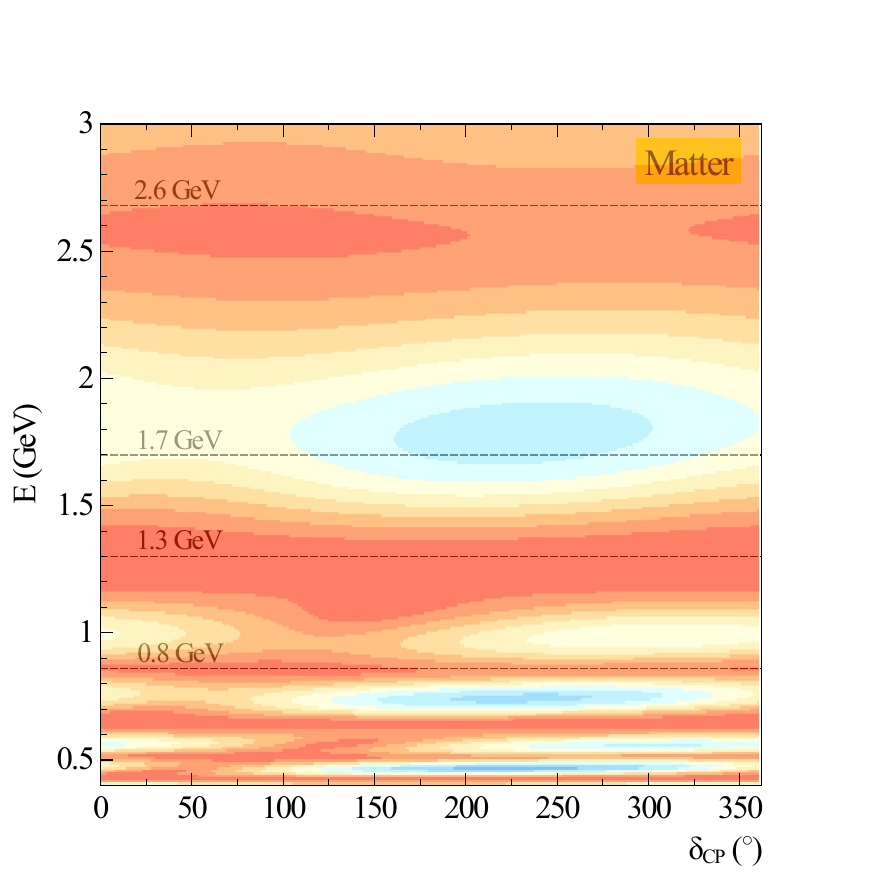}
\hspace{-2.38cm}
\includegraphics[scale=0.55]{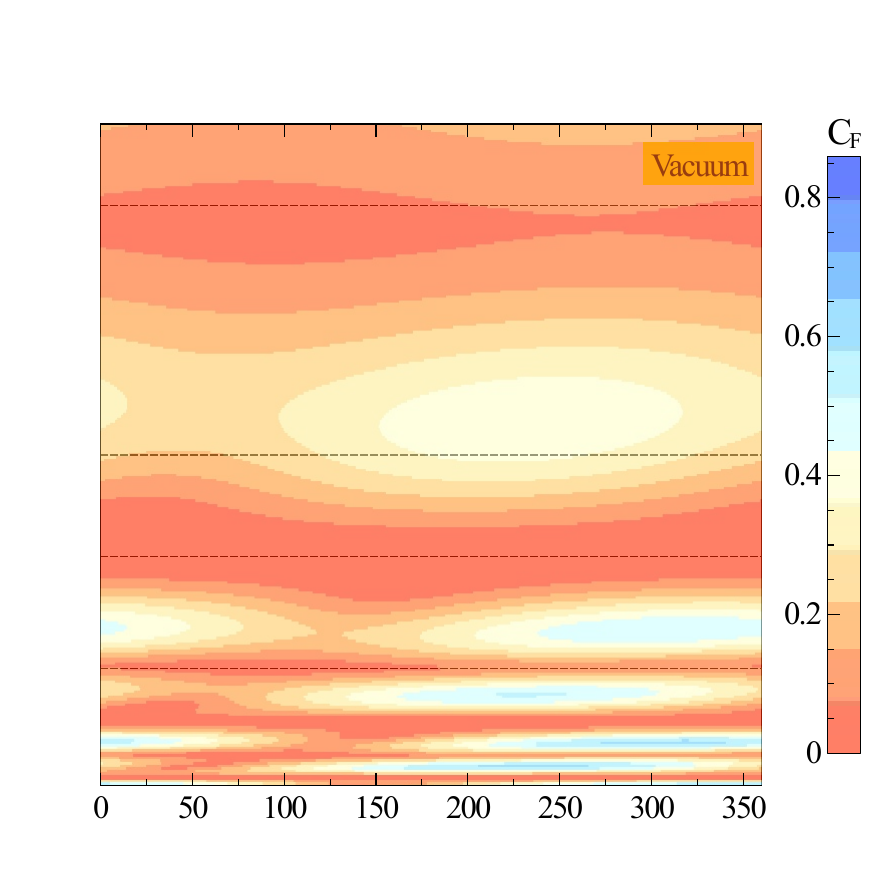}
\caption{ Oscillograms of the concurrence fill $C_F$ as functions of neutrino energy $E$ and the CP phase $\delta_{\rm CP}$ for a $1300~\mathrm{km}$ baseline, shown for matter (left panel) and for vacuum (right panel). The dashed horizontal lines mark representative energies corresponding to the first ($\sim 2.6~\mathrm{GeV}$) and second ($\sim 0.8~\mathrm{GeV}$) oscillation maxima, along with intermediate benchmark points ($E\sim 1.3~\mathrm{GeV}, 1.7~\mathrm{GeV}$). The color scale indicates the magnitude of $C_F$. Comparing the two panels demonstrates that matter effects reshape the distribution of genuine tripartite entanglement across the energy spectrum, modifying both the location and the intensity of regions where $C_F$ is enhanced or suppressed.}
\label{oscillogram}
\end{figure}

It will be interesting to study the sensitivity of the genuine tripartite entanglement measure like concurrence fill $C_F$ to the Dirac CP phase $\delta_{CP}$ and compare it with the standard oscillation fit. It is also important to identify the kinematic regions where the concurrence fill is most sensitive to the $\delta_{CP}$. 
To further clarify the physical role of the concurrence fill $C_F$ in probing $\delta_{CP}$, we present oscillograms showing the variation of $C_F$ as a function of neutrino energy $E$ and $\delta_{CP}$ in Fig.~\ref{oscillogram}. In these oscillograms, the vertical axis represents the neutrino energy $E$, the horizontal axis corresponds to the CP-violating phase $\delta_{CP}$, and the color scale encodes the magnitude of the concurrence fill $C_F$. Each point in the plot, therefore, corresponds to a fully evolved three-flavor neutrino state at fixed baseline, with the color indicating the strength of W-type tripartite correlations among the flavor modes.  The vacuum panel illustrates purely kinematic interference among mass eigenstates, while the matter panel incorporates the additional modifications induced by the matter potential.

From the perspective of CP-phase sensitivity, the key quantity is the variation of $C_F$ with respect to $\delta_{CP}$ i.e., the magnitude of the gradient, $\left| \frac{\partial C_F}{\partial \delta_{CP}} \right|$ which geometrically corresponds to the horizontal color variation in the oscillogram. The regions where the color changes rapidly along the $\delta_{CP}$ direction indicate strong phase responsiveness, meaning that small shifts in $\delta_{CP}$ produce appreciable changes in the entanglement structure. In contrast, horizontally uniform bands correspond to $\partial C_F/\partial \delta_{CP} \approx 0$, where the observable $C_F$ becomes effectively insensitive to the CP phase.

Next, we examine, in this context, three interesting kinematic regions corresponding to the energies $E\simeq 1.3$~GeV (minimal mixing and minimal $C_F$); $E\simeq 1.7$~GeV (maximal mixing and maximal $C_F$), and $E\simeq 2.6 (0.8)$~GeV for first (2nd) oscillation maxima with intermediate values for $C_F$. As may be observed in Fig.~\ref{oscillogram}, near $E \simeq 1.7~\mathrm{GeV}$, the concurrence fill reaches near-maximal values, signaling strong tripartite entanglement. This enhancement is not uniform in $\delta_{\mathrm{CP}}$, but is concentrated in the interval $\delta_{\mathrm{CP}} \approx 150^\circ\text{--}300^\circ$. At $E \simeq 1.3~\mathrm{GeV}$, the concurrence fill becomes strongly suppressed, and $C_F$ exhibits a nearly uniform horizontal band structure. Because the value of $C_F$ remains largely constant regardless of $\delta_{CP}$, the derivative approaches zero $\left( \frac{\partial C_F}{\partial \delta_{CP}} \approx 0 \right)$. Hence, in this regime, the physical state is effectively insensitive to the CP phase. 
At energies corresponding to the first and second oscillation maxima 
(e.g., $E \sim 2.6~\mathrm{GeV}$ and $0.8~\mathrm{GeV}$), the entanglement measure $C_F$ assumes a moderate values but exhibits pronounced horizontal gradients. In these regions, small variations in $\delta_{CP}$ produce significant changes in $C_F$, indicating enhanced phase sensitivity, thereby making  $C_F$ highly phase-dependent. 

In particular, the regions near the oscillation maxima --- especially around the second oscillation maximum where matter effects are comparatively moderate --- exhibit pronounced horizontal gradients, identifying favorable energy windows for enhanced $\delta_{CP}$ resolution. Thus, such an oscillogram serves as a useful guide for identifying optimal operating kinematic regimes in long-baseline experiments. 
However, when the primary objective is the precision determination of $\delta_{CP}$, the experimental focus should shift from regions of maximal $C_F$ to those where the observable exhibits the strongest phase sensitivity. Hence, considering $C_{F}$ as a potential CP-sensitivity observable, this moderately entangled region at $E\sim 0.8\,\mathrm{GeV}$, provides the most favorable conditions for precision measurement of $\delta_{CP}$.

\section{Conclusion}
\label{sec:conclusion}
In the present study, we have thoroughly examined the entangled structure of three-flavor neutrino oscillations within a three-qubit framework, highlighting that a single neutrino, although produced in a definite flavor state, naturally evolves into a coherent excitation distributed across three orthogonal flavor modes. This viewpoint demonstrates that neutrino oscillations encompass not only quantum superposition among mass eigen states but also constitute a genuine single-particle multi-mode entanglement phenomenon in the flavor basis.

We developed a comprehensive formalism to quantify this multi-mode structure through three complementary tripartite entanglement measures: (i) the tangle, (ii) the partial tangle, and (iii) the concurrence fill. Notably, each of these entanglement measures is expressed in terms of experimentally accessible appearance and disappearance probabilities. Using the density-matrix formalism, including matter effects, we derived analytic expressions for bipartite and bipartitioned concurrences, partial tangles, the three-tangle, and the concurrence fill. The ability to express all entanglement measures directly in terms of observable oscillation probabilities provides a bridge between quantum information theory and long baseline neutrino experiments. In particular, experiments such as DUNE, with broad energy coverage and high precision, provide an excellent platform for probing the dynamical behavior and distribution of multi-mode entanglement in the neutrino sector.

The DUNE experiment provides an exceptionally favorable setting for quantum informational studies of neutrino oscillations, owing to its broad band neutrino energy spectrum and substantial flux across the $0.5-3~\mathrm{GeV}$ range. This wide coverage allows DUNE to probe not only the first oscillation maximum but also the second, where the oscillation amplitude of $P_{\mu e}$ is amplified by a factor of three as compared to the first oscillation. The ability to access both oscillation maxima is crucial for resolving the dynamical redistribution of flavor mode entanglement in ways that narrow band beams cannot achieve.
Motivated by these advantages, the present work adopted the DUNE experimental configuration as the primary framework for our quantum informational analysis. All numerical results have been obtained from detailed simulations using the \textsf{GLoBES} software package, ensuring that the behavior of the entanglement measures accurately reflects realistic long-baseline oscillation dynamics.

Our results reveal a striking feature that the three-tangle vanishes identically throughout the entire dynamical evolution. Such an observation confirms that the tripartite entanglement of the three-flavor neutrino system is not GHZ-type, but rather lies within the class of separable, biseparable, or W-type configurations. Motivated by this, we analyzed the distribution of bipartite entanglement using the concurrences. It is observed that the entanglement is predominantly shared between two flavor modes at a given time. In particular, $C_{\mu|e\tau}$ and $C_{\tau|e\mu}$ exhibit nearly identical behavior across energy, whereas $C_{e|\mu\tau}$ becomes dominant at the oscillation maxima of $\nu_e$. While examining individual two-flavor concurrences, the maximal entanglement exists between $(\nu_\mu,\nu_\tau)$ for an initial $\nu_\mu$ state. In the absence of a $\nu_\mu$ component, the entanglement is primarily shared between $(\nu_e,\nu_\tau)$ while correlations between $(\nu_\mu,\nu_e)$ remain negligible. These observations demonstrate that bipartite entanglement follows the oscillatory pattern of the flavor transition probabilities. 

The partial-tangle analysis further clarifies the monogamy structure of neutrino entanglement. The three partial tangles exhibit clear oscillatory and anti-correlated patterns: close to the first oscillation maximum, one partial tangle, $\tau_{e\tau}$, becomes nearly maximal while the others, $\tau_{\mu\tau}$, $\tau_{\mu e}$, are strongly suppressed. This behavior illustrates explicit monogamy of entanglement, i.e., the enhancement of correlations in one pair ($\tau_{e\tau}$) leads to a diminishing of correlations in the remaining pairs ($\tau_{\mu\tau}$, $\tau_{\mu e}$). The redistribution of partial tangles preserves the topology of W-type entanglement with vanishing tangle for each flavor. 

We then computed the concurrence fill $C_F$ which is a geometric measure of genuine tripartite entanglement for the three flavors neutrino oscillations. The concurrence fill reflects the area enclosed by the concurrence triangle, with each side representing the bi-partitioned concurrence $C_{\alpha|\beta\gamma}$. The estimated area of the concurrence triangle varies with energy, reaching a minimum. The non-vanishing values for $C_F$ over most energies manifest the presence of W-type entanglement during the flavor evolution.
Further, W-class inequality provides a decisive classification of the entanglement structure. It is observed that, at $E = 1.3$~GeV, all the bipartite and tripartite entanglement measures vanish identically and the neutrino system behaves completely separable.  

We have also studied the sensitivity of the genuine tripartite entanglement measure, like concurrence fill $C_F$, to the Dirac CP phase $\delta_{CP}$ by the gradient $\left| \partial C_F / \partial \delta_{CP} \right|$ for different energies within the DUNE setup. In particular, we have examined three interesting cases corresponding to the energies $E\simeq 1.3$~GeV (minimal mixing and minimal $C_F$); $E\simeq 1.7$~GeV (maximal mixing and maximal $C_F$), and $E\simeq 2.6 (0.8)$~GeV for first (2nd) oscillation maxima and intermediate values for $C_F$. It turns out that at $E \simeq 1.3~\mathrm{GeV}$, $C_F$ is almost independent of $\delta_{\mathrm{CP}}$, and the gradient $\left| \partial C_F / \partial \delta_{CP} \right|$ is nearly zero. In this region, the tripartite entanglement measure $C_F$ is effectively blind to the CP phase. Analogously, at $E \simeq 1.7~\mathrm{GeV}$, the concurrence fill approaches its maximum, indicating strong tripartite entanglement, predominantly within $\delta_{\mathrm{CP}} \approx 150^\circ\text{--}300^\circ$. In contrast, near the first (e.g., $E \sim 2.6~\mathrm{GeV}$) and especially the second oscillation maxima ( $E \sim 0.8~\mathrm{GeV}$), the concurrence fill $C_F$ takes moderate values while the small variation in $\delta_{\mathrm{CP}}$ produces significant changes in $C_F$. Thus, the gradient in $C_F$ with respect to $\delta_{\rm CP}$ might play an important role in long baseline experiments like DUNE, identifying the interesting kinematic regions.

In summary, the present analysis demonstrates that the dynamical evolution of a three-flavor neutrino system provides a natural realization of single-particle multimode entanglement with its structure fully expressible in terms of experimentally measurable flavor transition oscillation probabilities. The framework developed here provides a unified method for identifying separable, biseparable, and W-type configurations as functions of energy and baseline, thereby establishing a direct bridge between quantum information diagnostics and long baseline neutrino phenomenology. It is important to investigate how these entanglement measures behave in the presence of nonstandard interactions, decay, or decoherence effects \cite{Dixit:2020ize,Chattopadhyay:2021eba}, which can impact the precision of measurements of oscillation parameters. The present formalism can be naturally extended to scenarios with sterile neutrinos, nonunitary mixing, or astrophysical events, i.e., supernova scenarios \cite{Dighe:1999bi,Das:2017iuj, Dighe:2017sur, Sen:2024fxa}.  
Thus, with the next generation facilities such as DUNE, Hyper-K, and ESSnuSB entering high precision regimes, the entanglement-based tools provided here offer a promising avenue for probing quantum-information aspects of neutrino flavor evolution in experimentally accessible settings.
\section*{Acknowledgments}
Rajrupa Banerjee would like to thank the Ministry of Electronics and IT for the financial support through the Visvesvaraya fellowship scheme for carrying out this research work. SP would like to acknowledge financial support from MTR/2023/000687, funded by SERB, Govt. of India. RB likes to acknowledge the Institute of Physics, Bhubaneswar, for its kind hospitality during the initial stages of this work. 

\bibliographystyle{jhep}
\bibliography{entanglement_neutrino}
\end{document}